\documentclass[runningheads]{llncs}

\usepackage{graphicx}
\graphicspath{{images/}}

\usepackage{tcolorbox}
\usepackage{verbatim}
\usepackage{multirow}
\usepackage{paralist}
\usepackage{enumitem}

\usepackage{amsmath}
\usepackage{amssymb}
\usepackage{wrapfig}

\usepackage{algorithm}
\usepackage{algorithmicx}
\usepackage{algpseudocode}

\usepackage[normalem]{ulem}

\usepackage{tikz}
\usepackage{listings,color}
\usetikzlibrary{shadows,shapes,shapes.geometric,arrows}
\usepackage{pgfplots}
\pgfplotsset{compat=1.10}

\newcommand*\circled[1]{\tikz[baseline=(char.base)]{
            \node[shape=circle,draw,inner sep=1.2pt] (char) {#1};}}

\usepackage{colortbl}
\definecolor{TableRowGray}{gray}{0.93}

\definecolor{dkgreen}{rgb}{0,0.6,0}
\definecolor{forestgreen}{RGB}{0,100,50}
\definecolor{redd}{RGB}{76,0,153}
\definecolor{english}{rgb}{0.0, 0.26, 0.15}
\definecolor{coolblack}{rgb}{0.0, 0.18, 0.39}
\definecolor{blue-violet}{rgb}{0.54, 0.17, 0.89}

\lstdefinestyle{customjava}
{language=java,
keywordstyle=\color{blue-violet},
basicstyle=\ttfamily\scriptsize,
morekeywords={var, require, msg, keccak256, now, function, continue, else, contract, enum, struct},
mathescape=true,
escapeinside={/*@}{@*/},
commentstyle=\color{english},   
keywordstyle=[2]{\color{red}},
frame=single
}

\lstdefinestyle{custombip}
{language=java,
keywordstyle=\color{blue},
basicstyle=\ttfamily\footnotesize,
morekeywords={package, atom, data, initial, place, on, from, provided, do, to, end},
mathescape=true,
escapeinside={/*@}{@*/},
commentstyle=\color{english},   
keywordstyle=[2]{\color{red}},
frame=single
}

\usepackage[textsize=tiny,textwidth=4.2cm,disable]{todonotes}

\newcommand{\code}[1]{\textcolor{blue}{\texttt{\small{#1}}}}
\newcommand{\lbl}[1]{\textit{#1}}
\newcommand{\ind}{\hspace{1em}}

\newcommand{\Natassa}[1]{\todo[color=red!10,linecolor=black!50]{\textbf{Natassa}: #1}}
\newcommand{\Aron}[1]{\todo[color=yellow!5,linecolor=black!50]{\textbf{Aron}: #1}}
\newcommand{\Emmanouela}[1]{\todo[color=green!5,linecolor=black!50]{\textbf{Emmanouela}: #1}}

\newcommand{\Stmt}[0]{\ensuremath{\textnormal{Stmt}}}
\newcommand{\Exp}[0]{\ensuremath{\textnormal{Exp}}}

\newcommand{\Type}[0]{\ensuremath{\textnormal{Type}}}

\newcommand{\True}[0]{\ensuremath{\texttt{true}}}
\newcommand{\False}[0]{\ensuremath{\texttt{false}}}

\newcommand{\Params}[2]{\textit{Params}(#1, #2)}
\newcommand{\Eval}[2]{\ensuremath{\textnormal{Eval}({#1}, {#2})}}
\newcommand{\Decl}[2]{\ensuremath{\textnormal{Decl}(#1, #2)}}
\newcommand{\Log}[0]{\ensuremath{\textnormal{Log}}}

\newcommand{\OSRule}[4]{{\noindent\begin{tabular}{cc}&$\begin{array}{c} #1 \end{array}$\\\cline{2-2}\multirow{-2}{*}{\small{#4}} \, & $\left\langle {#2} \right\rangle \rightarrow \left\langle {#3} \right\rangle$ \end{tabular} } }
\newcommand{\OSRuleBreak}[4]{{\noindent\begin{tabular}{cc}&$\begin{array}{c} #1 \end{array}$\\\cline{2-2}\multirow{-2}{*}{\small{#4}} \, & $\begin{array}{l}\left\langle {#2} \right\rangle \\ \rightarrow \left\langle {#3} \right\rangle\end{array}$ \end{tabular} } }

\usepackage[breaklinks,hidelinks]{hyperref}

\begin{document}


\title{VeriSolid:
Correct-by-Design\\Smart Contracts for Ethereum
}

\author{Anastasia Mavridou\inst{1} \and Aron Laszka\inst{2}\and\\  Emmanouela Stachtiari\inst{3} \and Abhishek Dubey\inst{1}}
\authorrunning{Mavridou et al.}

\institute{Vanderbilt University \and University of Houston \and Aristotle University of Thessaloniki}

\let\oldaddcontentsline\addcontentsline
\def\addcontentsline#1#2#3{}
\maketitle
\def\addcontentsline#1#2#3{\oldaddcontentsline{#1}{#2}{#3}}

\begin{center}
Accepted for publication in the proceedings of the\\23rd International Conference on Financial Cryptography and Data Security (FC 2019).
\end{center}

\begin{abstract}
The adoption of blockchain based distributed ledgers is growing fast due to their ability to provide reliability, integrity, and auditability without trusted entities.
One of the key capabilities of these emerging platforms is the ability to create self-enforcing smart contracts.
However, the development of smart contracts has proven to be error-prone in practice, and as a result, contracts deployed on public platforms are often riddled with security vulnerabilities.
This issue is exacerbated by the design of these platforms, which forbids updating contract code 
and rolling back malicious transactions.
In light of this, it is crucial to ensure that a smart contract is secure before deploying it and trusting it with significant amounts of cryptocurrency.
To this end, we introduce the \emph{VeriSolid} framework for the formal verification of contracts that are specified using a transition-system based model with rigorous operational semantics.  Our model-based approach allows developers to reason about and verify contract behavior at a high level of abstraction.  
VeriSolid allows the generation of Solidity code from the verified models, which enables the \emph{correct-by-design} development of smart~contracts. 
\end{abstract}
\vspace{-1.8em}


\newpage
\setcounter{tocdepth}{2}
\tableofcontents

\newpage
\section{Introduction}
\vspace{-0.3em}

The adoption of blockchain based platforms is rising rapidly.
Their popularity is explained by their ability to maintain a \emph{distributed public ledger}, providing reliability, integrity, and auditability \emph{without a trusted entity}.
Early blockchain platforms, e.g., Bitcoin, focused solely on creating cryptocurrencies and payment systems.
However, more recent platforms, e.g., Ethereum, also act as distributed computing platforms~\cite{underwood2016blockchain,wood2014ethereum} and enable the creation of \emph{smart contracts}, i.e., software code that runs on the platform and automatically executes and enforces the terms of a contract~\cite{clack2016smart}.
Since smart contracts can perform any computation\footnote{While the virtual machine executing a contract may be Turing-complete, the amount of computation that it can perform is actually limited in practice.}, they allow the development of decentralized applications, whose execution is safeguarded by the security properties of the underlying platform.
Due to their unique advantages, blockchain based platforms are envisioned to have a wide range of applications, ranging from financial to the Internet-of-Things~\cite{christidis2016blockchains}.

However, the trustworthiness of the platform guarantees only that a smart contract is executed correctly, not that the code of the contract is correct.
In fact, a large number of contracts deployed in practice suffer from software vulnerabilities, which are often introduced due to the semantic gap between the assumptions that contract writers make about the underlying execution semantics and the actual semantics of smart contracts~\cite{luu2016making}.
A recent automated analysis of 19,336 contracts deployed on the public Ethereum blockchain found that 8,333 contracts suffered from at least one security issue~\cite{luu2016making}.
While not all of these issues lead to security vulnerabilities, many of them enable stealing digital assets, such as cryptocurrencies.
Smart-contract vulnerabilities have resulted in serious security incidents, such as the ``DAO attack,'' in which \$50 million worth of cryptocurrency was stolen~\cite{finley2016million}, 
and the 2017 hack of the multisignature Parity Wallet library~\cite{newman2017security}, which lost \$280 million worth of cryptocurrency. 

The risk posed by smart-contract vulnerabilities is exacerbated by the typical design of blockchain based platforms, which does not allow the code of a contract to be updated (e.g., to fix a vulnerability) or a malicious transaction to be reverted.
Developers may circumvent the immutability of code by separating the ``backend'' code of a contract into a library contract that is referenced and used by a ``frontend'' contract, and updating the backend code by deploying a new instance of the library and updating the reference held by the frontend.
However, the mutability of contract terms introduces security and trust issues (e.g., there might be no guarantee that a mutable contract will enforce any of its original terms).
In extreme circumstances, it is also possible to revert a transaction by performing a hard fork of the blockchain.
However, a hard fork requires consensus among the stakeholders of the entire platform, undermines the trustworthiness of the entire platform, and may introduce security issues (e.g., replay attacks between the original and forked chains).

In light of this, it is crucial to ensure that a smart contract is secure before deploying it and trusting it with significant amounts of cryptocurrency.
%
Three main 
approaches have been considered for securing smart contracts, including
secure programming practices and patterns (e.g., Checks--Effects--Interactions pattern~\cite{solidityCEIpattern}), automated vulnerability-discovery tools (e.g., \textsc{Oyente}~\cite{luu2016making,tsankov2018securify}), and formal verification of correctness (e.g.,~\cite{hirai2017defining,grishchenko2018semantic}). 
Following secure programming practices and using common patterns can decrease the occurrence of vulnerabilities.
However, their effectiveness is limited for multiple reasons. 
First, they rely on a programmer following and implementing them, which is error prone due to human nature.
Second, they can prevent a set of typical vulnerabilities, but they are not effective against vulnerabilities that are atypical or belong to types which have not been identified yet. Third, they cannot provide formal security and safety guarantees.
%
Similarly, automated vulnerability-discovery tools consider generic properties that usually do not capture contract-specific requirements and thus, are effective in detecting typical errors but ineffective in detecting atypical vulnerabilities. These tools typically require security properties and patterns to be specified at a low level (usually bytecode) by security experts. Additionally, automated vulnerability-discovery tools are not precise; they often produce false positives.

On the contrary, formal verification tools are based on formal operational semantics and provide strong verification guarantees. They enable the formal specification and verification of properties and can detect both typical and atypical vulnerabilities that could lead to the violation of some security property. However, these tools are harder to automate.
%
%

Our approach falls in the category of formal verification tools, but it also provides an end-to-end design framework, 
 which combined with a code generator, allows the \emph{correctness-by-design} development of Ethereum smart contracts. We focus on providing usable tools for helping developers to eliminate errors early at design time by raising the abstraction level and employing graphical representations. Our approach does not produce false positives for safety properties and deadlock-freedom. 

In principle, a contract vulnerability is a programming error that enables an attacker to use a contract in a way that was not intended by the developer.
To detect vulnerabilities that do not fall into 
common types, developers must specify the intended behavior of a contract.
Our framework enables developers to specify intended behavior 
in the form of liveness, deadlock-freedom, and safety properties, which capture important security concerns and vulnerabilities.
One of the key advantages of our model-based verification approach is that it allows developers to specify desired properties with respect to high-level models instead of, e.g., bytecode.
%
Our tool can then automatically verify whether the behavior of the contract satisfies these properties.
If a contract does not satisfy some of these properties, our tool notifies the developers, explaining the execution sequence that leads to the property violation.
The sequence can help the developer to identify and correct the design errors that lead to the erroneous behavior. Since the verification output provides guarantees to the developer regarding the actual execution semantics of the contract, it helps eliminating the semantic gap.
%
Additionally, our verification and code generation approach fits smart contracts well because contract code cannot be updated after deployment. Thus, code generation needs to be performed only once before deployment. 

\vspace{-0.3cm}
\subsubsection*{Contributions}
We build on the \emph{FSolidM}~\cite{mavridou2018designing,mavridou2018tool} framework, which provides a graphical editor for specifying Ethereum smart contracts as transitions systems and a \emph{Solidity} code generator.\footnote{Solidity is  the high-level language for developing Ethereum contracts. Solidity code can be compiled into bytecode, which can be executed on the Ethereum platform.}
We present the \emph{VeriSolid} framework, which 
introduces \emph{formal verification capabilities}, thereby providing an approach for correct-by-design development of smart contracts. Our contributions are: 
\begin{itemize}
\item We extend the syntax of FSolidM models (Definition~\ref{def:smartContract}), provide formal operational semantics 
(FSolidM has no formal operational semantics) for our model (Section~\ref{sec:semantics}) and for supported Solidity statements (Appendix~\ref{app:soliditySemantics}), and extend the Solidity code generator (Appendix~\ref{sec:transformation}).
\item We design and implement developer-friendly natural-language like templates for specifying safety and liveness properties (Section~\ref{sec:templates}).
\Aron{Maybe state that we refer to the extended versions as VeriSolid to avoid the CCS'18 comment about self-plagarism.}
\item The developer input of VeriSolid is a transition system, in which each transition action  is specified using Solidity code. We provide an automatic transformation from the initial system into an augmented transition system, which extends the initial system with the control flow of the Solidity action of each transition (Section~\ref{sec:augmented}). We prove that the initial and augmented transition systems are observationally equivalent (Section~\ref{sec:proof}); thus, the verified properties of the augmented model are also guaranteed in the initial model.
\item We use an overapproximation approach for the meaningful and efficient verification of smart-contract models (Section~\ref{sec:verificationProcess}). We integrate verification tools (i.e., nuXmv and BIP) and present verification results. 
\end{itemize}

 \section{VeriSolid: Design and Verification WorkFlow}
\label{sec:workflow}

\begin{figure}[h!]
\centering
\includegraphics[scale=0.9]{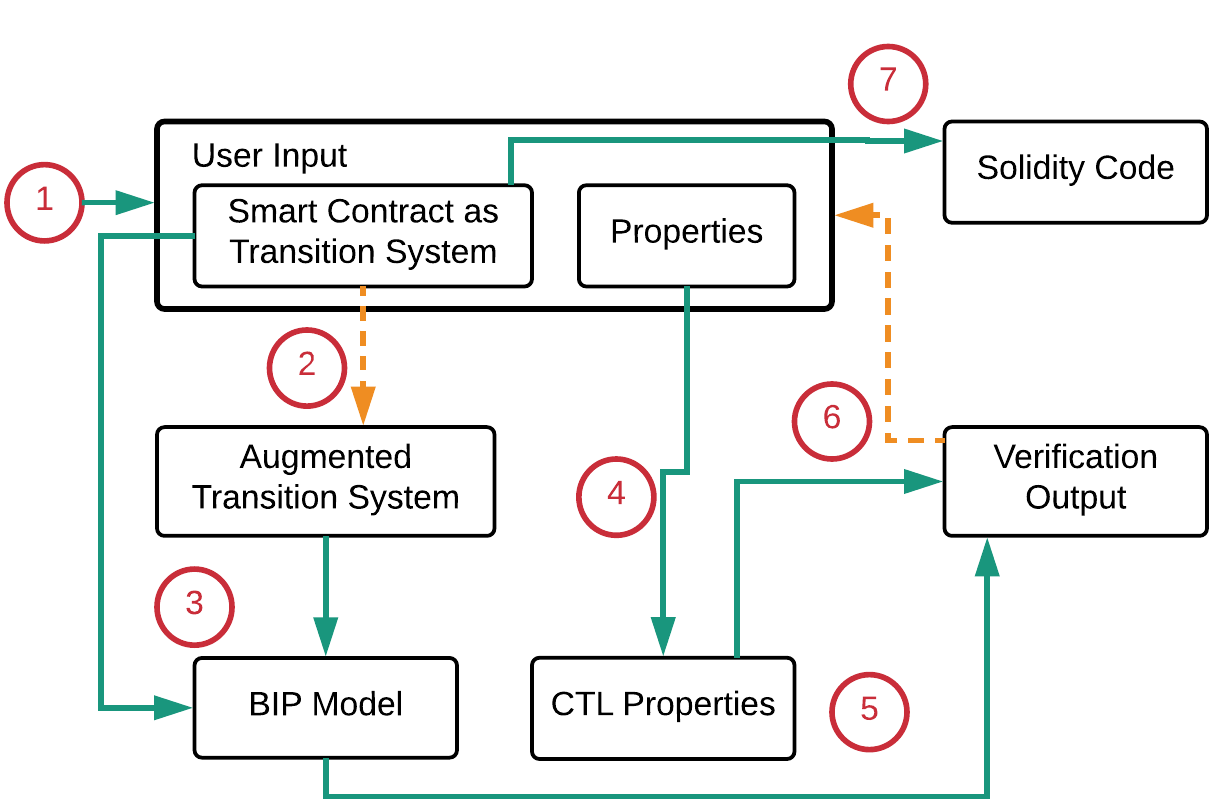}
\caption{Design and verification workflow.}
\label{fig:flow}
\end{figure}

\begin{figure}[h!]
    \centering
    \includegraphics[width=0.9\textwidth]{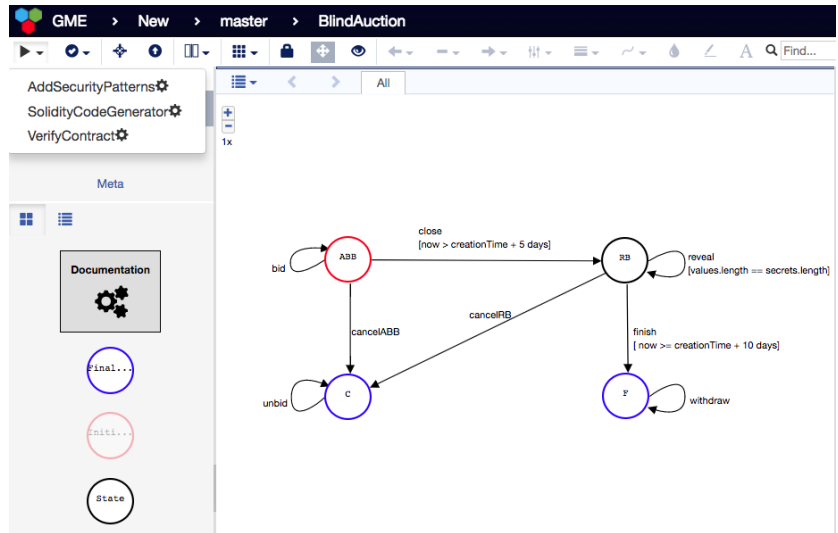}
    \caption{WebGME based graphical editor.}
    \label{fig:editor}
\end{figure}

%
VeriSolid is an open-source\footnote{\url{https://github.com/anmavrid/smart-contracts}} and web-based framework that is built on top of \mbox{WebGME}~\cite{maroti2014next} and FSolidM~\cite{mavridou2018designing,mavridou2018tool}. VeriSolid allows the collaborative development of Ethereum contracts with built-in version control, which enables branching, merging, and history viewing.
Figure~\ref{fig:flow} shows the steps of the VeriSolid design flow. Mandatory steps are represented by solid arrows, while optional steps are represented by dashed arrows. 
In step \circled{1}, the developer input is given, which consists of:
\begin{itemize}[topsep=0pt, nosep]
\item A contract specification containing 1)~a graphically specified transition system and 2) variable declarations, actions, and guards specified in Solidity.
\item A list of properties to be verified, which can be expressed using predefined natural-language like templates.
\end{itemize}
Figure~\ref{fig:editor} shows the web-based graphical editor of VeriSolid. 

The verification loop starts at the next step.  Optionally, step \circled{2} is automatically executed if the verification of the specified properties requires the generation of an augmented contract model\footnote{We give the definition of an augmented smart contract in Section~\ref{sec:augmented}.}.  
Next, in step \circled{3}, the Behavior-Interaction-Priority (BIP) model of the contract (augmented or not) is automatically generated. Similarly, in step \circled{4}, the specified properties are automatically translated to Computational Tree Logic (CTL). The model can then be verified for deadlock freedom or other properties using tools from the BIP tool-chain~\cite{basu2011rigorous} or nuXmv~\cite{bliudze2015formal} (step \circled{5}). If the required properties are not satisfied by the model (depending on the output of the verification tools), the specification can be refined by the developer (step \circled{6}) and analyzed anew. Finally, when the developers are satisfied with the design, i.e., all specified properties are satisfied, the equivalent Solidity code of the contract is automatically generated in step \circled{7}. The following sections describe the steps from Figure~\ref{fig:flow} in detail. Due to space limitations, we present the Solidity code generation (step \circled{7}) in Appendix~\ref{sec:transformation}.

 \section{Developer Input: Transition Systems and Properties}
\label{sec:designInput}


\subsection{Smart Contracts as Transition Systems}

To illustrate how to represent smart contracts as transition systems, we use the \emph{Blind Auction} example from prior work~\cite{mavridou2018designing}, which is based on an example from the Solidity documentation~\cite{solidityExample}.

In a blind auction, each bidder first makes a deposit and submits a blinded bid, which is a hash of its actual bid, and then reveals its actual bid after all bidders have committed to their bids.
After revealing, each bid is considered valid if it is higher than the accompanying deposit, and the bidder with the highest valid bid is declared winner.
A blind auction contract has four main~states:
%
\begin{enumerate}[topsep=0pt, nosep]
\item \texttt{AcceptingBlindedBids}: 
bidders submit blinded bids and make deposits;
\item \texttt{RevealingBids}: 
bidders reveal their actual bids by submitting them to the contract, and the contract checks for each bid that its hash is equal to the blinded bid and that it is less than or equal to the deposit made earlier;
\item \texttt{Finished}: winning bidder (i.e., the bidder with the highest valid bid) withdraws the difference between her deposit and her bid; other bidders withdraw their entire deposits;
\item \texttt{Canceled}: all bidders withdraw their deposits (without declaring a winner).
\end{enumerate}

\begin{figure}[h!]
\centering
\includegraphics[scale=0.95]{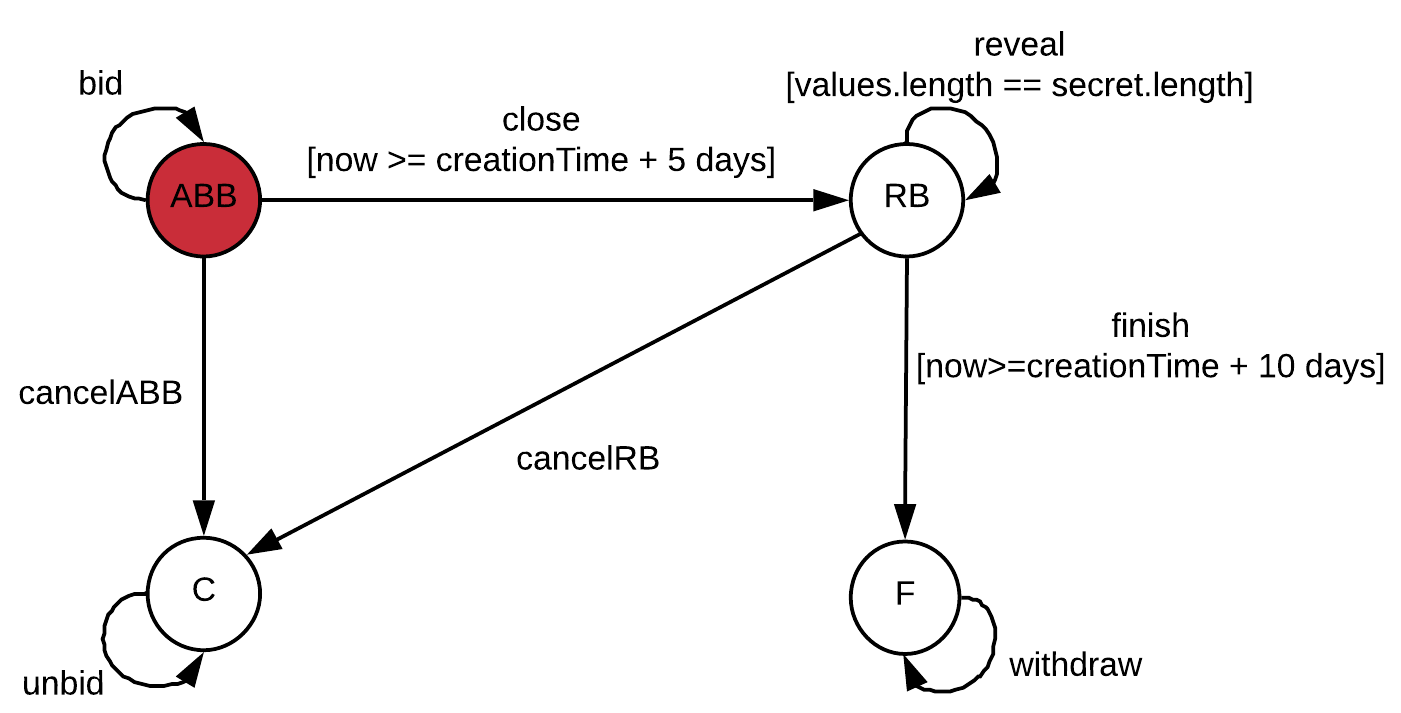}
\caption{Blind auction example as a transition system.}
\label{fig:blauction}
\end{figure}

This example illustrates that smart contracts have \emph{states} (e.g., \texttt{Finished}).
Further, contracts provide functions, which allow other entities (e.g., users or contracts) to invoke \emph{actions} and change the states of the contracts.
Hence, we can represent a smart contract naturally as a \emph{transition system}~\cite{solidityPatterns}, which comprises a set of states and a set of transitions between those states.
Invoking a transition forces the contract to execute the action of the transition if the \emph{guard} condition of the transition is satisfied.
Since such states and transitions have intuitive meanings for developers, representing contracts as transition systems provides an adequate level of abstraction for reasoning about their behavior.

Figure \ref{fig:blauction} shows the blind auction example in the form of a transition system.
For ease of presentation, we abbreviate \texttt{AcceptingBlindedBids}, \texttt{RevealingBids}, \texttt{Finished}, and \texttt{Canceled} to \texttt{ABB}, \texttt{RB}, \texttt{F}, and \texttt{C}, respectively.
The initial state of the transition system is \texttt{ABB}.
To differentiate between transition names and guards, we use square brackets for the latter.
Each transition (e.g., \texttt{close}, \texttt{withdraw}) corresponds to an action that a user may perform during the auction.
For example, a bidding user may execute transition \texttt{reveal} in state \texttt{RB} to reveal its blinded bid.
As another example, a user may execute transition \texttt{finish} in state \texttt{RB}, which ends the revealing phase and declares the winner, if the guard condition \texttt{now >= creationTime + 10 days} is true.
A user can submit a blinded bid using transition \texttt{bid}, close the bidding phase using transition \texttt{close}, and withdraw her deposit (minus her bid if she won) using transitions \texttt{unbid} and \texttt{withdraw}.
Finally, the user who created the auction may cancel it using transitions \texttt{cancelABB} and \texttt{cancelRB}.
For clarity of presentation, we omitted from Figure \ref{fig:blauction} the specific actions that the transitions take (e.g., transition \texttt{bid} executes---among others---the following statement: \texttt{pendingReturns[msg.sender] += msg.value;}).

\subsection{Formal Definition of a Smart Contract}

We formally define a contract as a transition system. To do that, we consider a subset of Solidity statements, which are described in detail in Appendix~\ref{app:soliditySubset}. 
We chose this subset of Solidity statements because it includes all the essential control structures: loops, selection, and \texttt{return} statements.  Thus, it is a Turing-complete subset, and can be extended in a straightforward manner to capture all other Solidity statements.
Our Solidity code notation is summarized in 
Table~\ref{tab:notation}. 

\begin{table}[h!]
\centering
\caption{Summary of Notation for Solidity Code}
\label{tab:notation}
\small
\begin{tabular}{|c|l|}
\hline
\small{Symbol}       & \multicolumn{1}{c|}{\small{Meaning}} \\ 
\hline
\small{$\mathbb{T}$} & \small{set of Solidity types} \\
\rowcolor{TableRowGray} \small{$\mathbb{I}$} & \small{set of valid Solidity identifiers} \\
\small{$\mathbb{D}$} & \small{set of Solidity event and custom-type definitions} \\
\rowcolor{TableRowGray} \small{$\mathbb{E}$} & \small{set of Solidity expressions} \\
\small{$\mathbb{C}$} & \small{set of Solidity expressions without side effects} \\
\rowcolor{TableRowGray} \small{$\mathbb{S}$} & \small{set of supported Solidity statements} \\ 
\hline
\end{tabular}
\end{table}

\newpage
\begin{definition}
\label{def:smartContract}
A transition-system \emph{initial smart contract} is a tuple $(D, S,$ $S_F,$ $s_0,$ $a_0, a_F, V, T)$, where 
\begin{compactitem}
\item $D \subset \mathbb{D}$ is a set of custom event and type definitions;
\item $S \subset \mathbb{I}$ is a finite set of states;
\item $S_F \subset S$ is a set of final states;
\item $s_0 \in S$, $a_0 \in \mathbb{S}$ are the initial state and action;
\item $a_F \in \mathbb{S}$ is the fallback action;
\item $V \subset \mathbb{I} \times \mathbb{T}$  contract variables (i.e., variable names and types);
\item $T \subset \mathbb{I} \times S \times 2^{\mathbb{I} \times \mathbb{T}} \times \mathbb{C} \times (\mathbb{T} \,\cup\, \emptyset) \times \mathbb{S} \times S$ is a transition relation, where each transition $\in T$ includes:
\begin{compactitem}
    \item transition name $t^\textit{name} \in \mathbb{I}$;
    \item source state $t^\textit{from} \in S$;
    \item parameter variables (i.e., arguments) $t^\textit{input} \subseteq {\mathbb{I} \times \mathbb{T}}$;
    \item transition guard $g_t \in \mathbb{C}$;
    \item return type $t^\textit{output} \in (\mathbb{T} \,\cup\, \emptyset)$;
    \item action $a_t \in \mathbb{S}$;
    \item destination state $t^\textit{to} \in S$.
\end{compactitem}
\end{compactitem}
\end{definition}

The initial action $a_0$ represents the constructor of the smart contract. A contract can have \textit{at most one constructor}. In the case that the initial action $a_0$ is empty (i.e., there is no constructor), $a_0$ may be omitted from the transition system. A constructor is graphically represented in VeriSolid as an incoming arrow to the initial state. 
The fallback action $a_F$ represents the fallback function of the contract.
Similar to the constructor, a contract can have \textit{at most one fallback} function. Solidity fallback functions are further discussed in Appendix~\ref{app:functionCalls}. 

\subsubsection{Lack of the Re-entrancy Vulnerability}
\label{sec:lessvulnerable}
\Aron{Too much focus on reentrancy vulnerability considering how simplistic our solution is. Incorporate CCS response!}
VeriSolid allows specifying contracts such that the re-entrancy vulnerability is prevented by design. In particular, after a transition begins but before the execution of the transition action, the contract changes its state to a temporary one (see Appendix~\ref{sec:transformation}). This prevents re-entrancy since none of the contract functions\footnote{Our framework implements transitions as  functions, see Appendix~\ref{sec:transformation}.} can be called in this state. 
One might question this design decision since re-entrancy is not always harmful. However, we consider that it can pose significant challenges for providing security. First, supporting re-entrancy substantially increases the complexity of verification. Our framework allows the efficient verification---within seconds---of a broad range of properties, which is essential for iterative development. Second, re-entrancy often leads to vulnerabilities since it significantly complicates contract behavior. We believe that prohibiting re-entrancy is a small price to pay for security. 

\subsection{Smart-Contract Operational Semantics}
\label{sec:semantics}

We define the operational semantics of our transition-system based smart contracts 
in the form of Structural Operational Semantics (SOS) rules~\cite{plotkin1981structural}. We let $\Psi$ denote the state of the ledger, which includes account balances, values of state variables in all contracts, number and timestamp of the last block, etc.
During the execution of a transition, the execution state $\sigma = \{\Psi, M\}$ also includes the memory and stack state~$M$. 
To handle return statements and exceptions, we also introduce an execution status, which is $E$ when an exception has been raised, $R[v]$ when a return statement has been executed with value $v$ (i.e., \texttt{return} $v$), and $N$ otherwise.
Finally, we let $\Eval{\sigma}{\Exp} \rightarrow \langle (\hat{\sigma}, x), v\rangle$ signify that the evaluation of a Solidity expression $\Exp$ in execution state~$\sigma$ yields value $v$ and---as a side effect---changes the execution state to~$\hat{\sigma}$ and the execution status to $x$.\footnote{Note that the correctness of our transformations does not depend on the exact semantics of $\textnormal{Eval}$.}

A transition is triggered by providing a transition (i.e., function) $\textit{name} \in \mathbb{I}$ and a list of parameter values $v_1, v_2, \ldots$.
The normal execution of a transition without returning any value, which takes the ledger from state $\Psi$ to  $\Psi'$ and the contract from state $s \in S$ to  $s' \in S$, is captured by the \textsc{TRANSITION} rule:

\vspace{0.5em}
\OSRule{
t \in T , \hspace{0.8em}  
\textit{name} = t^{name}  , \hspace{0.8em}
s = t^{\textit{from}}  \\
M = \Params{t}{v_1, v_2, \ldots} , \hspace{0.8em}
\sigma = (\Psi, M) \\
\Eval{\sigma}{g_t} \rightarrow \langle (\hat{\sigma}, N), \True \rangle \\
\langle (\hat{\sigma}, N), a_t \rangle \rightarrow \langle (\hat{\sigma}', N), \cdot \rangle \\
\hat{\sigma}' = (\Psi', M') , \hspace{0.8em}
s' = t^{\textit{to}}
}{
(\Psi, s), \textit{name}\,(v_1, v_2, \ldots)
}{
(\Psi', s', \cdot) 
}{
TRANSITION
}
\vspace{0.5em}

This rule is applied if there exists a transition $t$ whose name $t^{\textit{name}}$ is $\textit{name}$ and whose source state $t^{\textit{from}}$ is the current contract state $s$ (first line).
The execution state $\sigma$ is initialized by taking the parameter values $\Params{t}{v_1, v_2, \ldots}$ and the current ledger state $\Psi$ (second line).
If the guard condition $g_t$ evaluates  $\Eval{\sigma}{g_t}$ in the current state $\sigma$ to $\True$ (third line), then the action statement $a_t$ of the transition is executed (fourth line), which results in an updated execution state $\hat{\sigma}'$ (see statement rules in Appendix~\ref{app:soliditySemantics}).
Finally, if the resulting execution status is normal $N$ (i.e., no exception was thrown), then the updated ledger state~$\Psi'$ and updated contract state $s'$ (fifth line) are made permanent.

We also define SOS rules for all cases of erroneous transition execution (e.g., exception is raised during guard evaluation, transition is reverted, etc.) and for returning values. Due to space limitations, we include these rules in Appendix~\ref{app:transitionSemantics}.
We also define SOS rules for supported statements in Appendix~\ref{app:soliditySemantics}.

\subsection{Safety, Liveness, and Deadlock Freedom}
\label{sec:templates}

A VeriSolid model is automatically verified for deadlock freedom. A developer may additionally verify safety and liveness properties. To facilitate the specification of properties, VeriSolid offers a set of predefined natural-language like templates, which correspond to properties in CTL. Alternatively, properties can be specified directly in CTL. 
Let us go through some of these predefined templates. Due to space limitations, the full template list, as well as the CTL property correspondence is provided in Appendix~\ref{app:templates}. 

\begin{tcolorbox}[boxsep=2pt,left=0pt,right=0pt,top=0pt,bottom=0pt]
\label{eq:one}
\textbf{$\langle{\textit{Transitions}\,\cup\,\textit{Statements}\rangle}$} cannot happen after \textbf{$\langle{\textit{Transitions}\,\cup\,\textit{Statements}\rangle}$}.
\end{tcolorbox}
The above template expresses a safety property type. 
\textbf{Transitions} is a subset of the transitions of the model (i.e., $\textbf{Transitions} \subseteq T$). 
 A statement from \textbf{Statements} is a specific inner statement from the action of a specific transition (i.e., $\textbf{Statements} \subseteq\ T \times \mathbb{S}$).
%
For instance, we can specify the following safety properties for the Blind Auction example:
\begin{itemize}[topsep=0pt]
\item \textbf{bid} cannot happen after \textbf{close}.
\item \textbf{cancelABB; cancelRB} cannot happen after \textbf{finish},
\end{itemize}
where \textbf{cancelABB; cancelRB} means \textbf{cancelABB $\cup$ cancelRB}.

\begin{tcolorbox}[boxsep=2pt,left=0pt,right=0pt,top=0pt,bottom=0pt]
If \textbf{$\langle{\textit{Transitions}\,\cup\,\textit{Statements}\rangle}$} happens, \textbf{$\langle{\textit{Transitions}\,\cup\,\textit{Statements}\rangle}$} can happen only after \textbf{$\langle{\textit{Transitions}\,\cup\,\textit{Statements}\rangle}$} happens.
\end{tcolorbox}

The above template expresses a safety property type. 
%
A typical vulnerability is that currency withdrawal functions, e.g., \texttt{transfer}, allow an attacker to withdraw currency again before updating her balance (similar to ``The DAO'' attack). To check this vulnerability type for the Blind Auction example, we can specify the following property. The statements in the action of transition \texttt{withdraw} are shown in Figure~\ref{fig:withdraw}.
\begin{itemize}[topsep=0pt]
\item if \textbf{withdraw.\texttt{msg.sender.transfer(amount);}} happens, \newline \textbf{withdraw.\texttt{msg.sender.transfer(amount);}} can happen only after \newline \textbf{withdraw.\texttt{pendingReturns[msg.sender]=0;}} happens.
\end{itemize}
As shown in the example above, a statement is written in the following form: \textbf{\textit{Transition.Statement}} to refer to a statement of a specific transition. If there are multiple identical statements in the same transition, then all of them are checked for the same property. \Aron{AFAIK this is \textbf{incorrect}, our tools does not do this. Should we talk about how to handle multiple identical statements within a transition? CCS reviewers asked.}\Natassa{Yeah, probably thats a good idea.} To verify properties with statements, we need to transform the input model into an augmented model, as presented in Section \ref{sec:augmented}.

\begin{figure}[t]
\centering
\begin{lstlisting}[style=customjava] 
uint amount = pendingReturns[msg.sender];
if (amount > 0) {
  if (msg.sender!= highestBidder)
    msg.sender.transfer(amount);
  else
    msg.sender.transfer(amount - highestBid);
  pendingReturns[msg.sender] = 0;
}
\end{lstlisting}
\caption{Action of transition \texttt{withdraw} in Blind Auction, specified using Solidity.}
\label{fig:withdraw}
\end{figure}

\begin{tcolorbox}[boxsep=2pt,left=0pt,right=0pt,top=0pt,bottom=0pt]
\textbf{$\langle{\textit{Transitions}\,\cup\,\textit{Statements}\rangle}$} will eventually happen after \textbf{$\langle{\textit{Transitions}\,\cup\,\textit{Statements}\rangle}$}.
\end{tcolorbox}

Finally, the above template expresses a liveness property type. 
%
\label{ex:liveness}
For instance, with this template we can write the following liveness property for the Blind Auction example to check the Denial-of-Service vulnerability (Appendix~\ref{app:vulnerabilities}): 
\begin{itemize}[topsep=0pt]
\item \textbf{withdraw.\texttt{pendingReturns[msg.sender]=0;}} 
will eventually happen after  \textbf{withdraw.\texttt{msg.sender.transfer(amount);}}.
\end{itemize}

 \section{Augmented Transition System Transformation}
\label{sec:augmented}

To verify a model with Solidity actions, we transform it to a functionally equivalent model that can be input into our verification tools.
We perform two transformations:
First, we replace the initial action $a_0$ and the fallback action $a_F$ with transitions.
Second, we replace transitions that have complex statements as actions with a series of transitions that have only simple statements (i.e., variable declaration and expression statements).
After these two transformations, the entire behavior of the contract is captured using only transitions. The transformation algorithms are discussed in detail in Appendices~\ref{app:conformance} and~\ref{sec:augmentation}.
The input of the transformation is a smart contract defined as a transition system (see Definition~\ref{def:smartContract}). 
The output of the transformation is an \emph{augmented smart contract}:

\begin{definition}
An \emph{augmented contract} is a tuple $(D, S,$ $S_F,$ $s_0,$ $V, T)$, where
\begin{compactitem}
\item $D \subset \mathbb{D}$ is a set of custom event and type definitions;
\item $S \subset \mathbb{I}$ is a finite set of states;
\item $S_F \subset S$ is a set of final states;
\item $s_0 \in S$, is the initial state;
\item $V \subset \mathbb{I} \times \mathbb{T}$  contract variables (i.e., variable names and types);
\item $T \subset \mathbb{I} \times S \times 2^{\mathbb{I} \times \mathbb{T}} \times \mathbb{C} \times (\mathbb{T} \,\cup\, \emptyset) \times \mathbb{S} \times S$ is a transition relation (i.e., transition name, source state, parameter variables, guard, return type, action, and destination state).
\end{compactitem}
\end{definition}

\begin{figure}[h!]
 \centering
 \vspace{-2em}
 \includegraphics[scale=0.75]{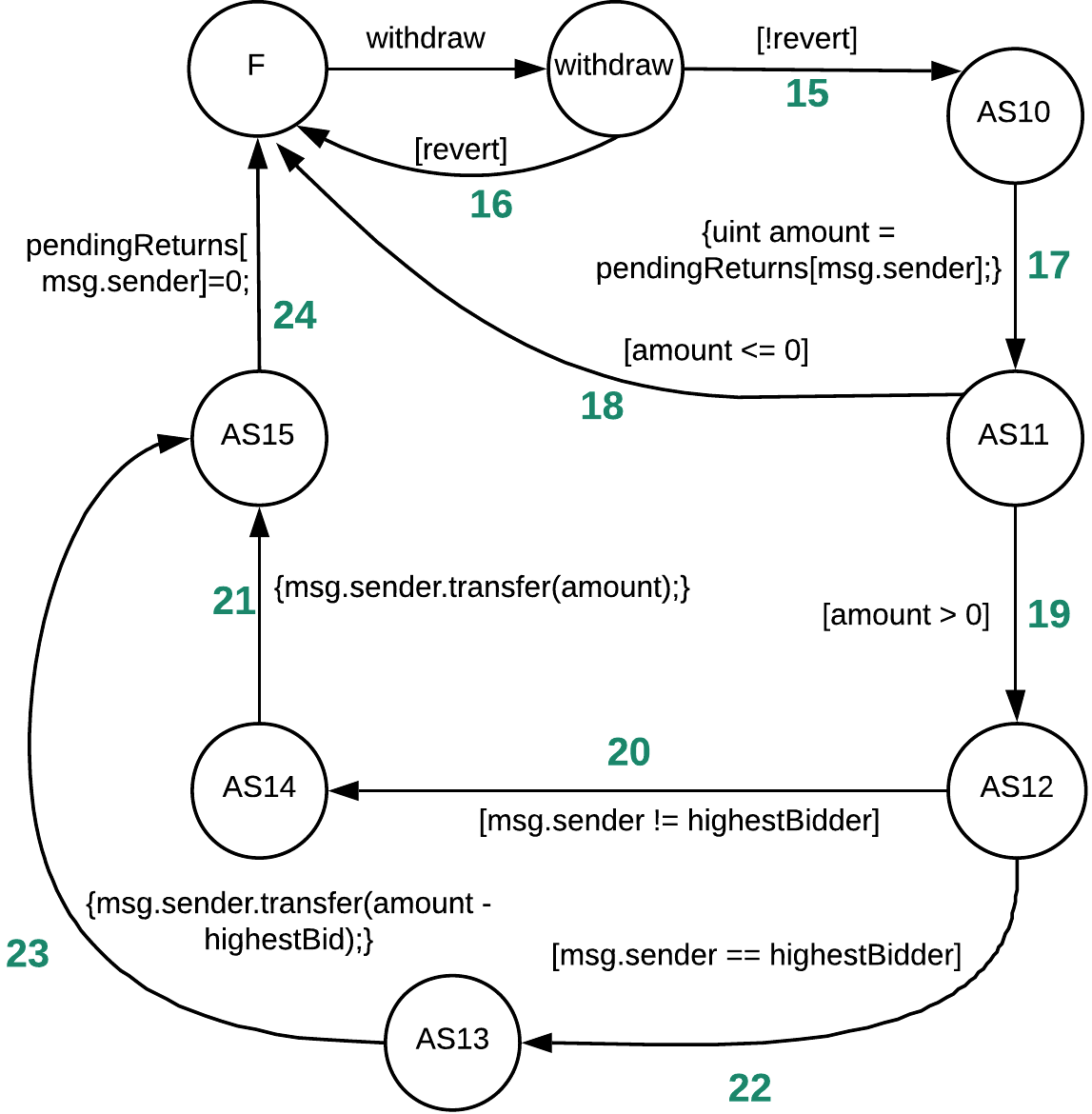}
 \vspace{-0.5em}
 \caption{Augmented model of transition \texttt{withdraw}.}
 \label{fig:withdrawModel}
  \vspace{-1.25em}
\end{figure}

Figure \ref{fig:withdrawModel} shows the augmented \texttt{withdraw} transition of the Blind Auction model. 
We present the complete augmented model in Appendix~\ref{sec:completeBlindAuction}. 
The action of the original \texttt{withdraw} transition is shown by Figure~\ref{fig:withdraw}.
Notice the 
added state \texttt{withdraw}, which \Aron{?} avoids re-entrancy by design, as explained in Section~\ref{sec:lessvulnerable}.

\subsection{Observational Equivalence}
\label{sec:proof}
We study sufficient conditions for augmented models to be behaviorally equivalent to initial models. To do that, we use observational equivalence~\cite{milner1989communication} by considering non-observable $\beta-$transitions. 
%
%
We denote by $S_{I}$ and $S_{E}$ the set of states of the smart contract transition system and its augmented derivative, respectively. We show that $R=\{(q,r) \in S_{I} \times S_{E} \}$
is a weak bi-simulation by considering 
as observable transitions $A$, those that affect the ledger state, while the remaining transitions $B$ are considered non-observable transitions. According to this definition, the set of transitions in the smart contract system, which represent the execution semantics of a Solidity named function or the fallback, are all observable. On the other hand, the augmented system represents each Solidity function using paths of multiple transitions. We assume that final transition of each such path is an $\alpha$ transition, while the rest are $\beta$ transitions.
Our weak bi-simulation is based on the fact the effect of each $\alpha \in A$ on the ledger state is equal for the states of $S_{I}$ and $S_{E}$. Therefore, if $\sigma_I=\sigma_E$ at the initial state of $\alpha$, then $\sigma'_I=\sigma'_E$ at the resulting state.

A weak simulation over $I$ and $E$ is a relation $R \subseteq S_{I} \times S_{E}$ such that we have:
\begin{description}

\item[Property 1] For all $(q,r) \in R$ and for each $\alpha \in A$, such that $q \overset{\alpha}{\rightarrow} q' $, there is $r'$ such that  $ r \overset{\beta^\star\alpha\beta^\star}{\rightarrow} r'$ where $(q',r') \in R $

For each observable transition $\alpha$ of a state in $S_{I}$, it should be proved that (i) a path that consists of $\alpha$ and other non-observable transitions exists in all its equivalent states in $S_{E}$, and (ii) the resulting states are equivalent.

\item[Property 2] For all $(q,r) \in R$ and $\alpha \in A$, such that $r \overset{\alpha}{\rightarrow} r' $, there is $q'$ such that $ q \overset{\alpha}{\rightarrow} q' $ where $(q',r') \in R $.
\\
For each observable outgoing transition in a state in $S_{E}$, it should be proved that (i) there is an outgoing observable transition in all its equivalent states in $S_{I}$, and (ii) the resulting states are equivalent.

\item[Property 3] 
 For all $(q,r) \in R$ and $\beta \in B$ such that $r\overset{\beta}{\rightarrow} r' $, $ (q,r') \in R$
\\ 
For each non observable transition, it should be proved that the the resulting state is equivalent with all the states that are equivalent with the initial state.

\end{description}

\begin{theorem}
\label{th:bisim}
For each initial smart contract $I$ and its corresponding augmented smart contract $E$, it holds that $I \sim E$.
\end{theorem}

The proof of Theorem~\ref{th:bisim} is presented in the Appendix~\ref{app:correctnessProof}.  

 \algrenewcommand{\algorithmiccomment}[1]{{\footnotesize[\textit{ #1 :}}}
\renewcommand{\alglinenumber}{\footnotesize}
\algblockdefx{StartAt}{EndAt}[1]{\textbf{atom type} #1}{\textbf{end}}
\algblockdefx{StartDir}{EndDir}[1]{\footnotesize[\textit{ #1 :}}{]}

\newcommand{\InComm}[3]{ {\footnotesize[\textit{ #1 :}} #2  {\footnotesize \textit{ #3} ]} }
\newcommand{\grt}[1]{\textgreater #1}
\newcommand{\lss}[1]{\textless #1}
\newcommand{\hlq}[1]{$\lss#1\grt$}
\newcommand{\noendtags}[0]{\algnotext{EndEach}} 
\newcommand{\emptylist}{\texttt{[]}}

\section{Verification Process}
\label{sec:verificationProcess}


Our verification approach checks whether contract behavior satisfies properties that are required by the developer. To check this, we must take into account the effect of data and time. However, smart contracts use environmental input as control data, e.g., in guards. Such input data can be infinite, leading to infinitely many possible contract states. Exploring every such state is highly inefficient~\cite{clarke1994abstract} and hence, appropriate data and time abstractions must be employed.

We apply data abstraction to ignore variables that depend on (e.g., are updated by) environmental input.
Thus, an overapproximation of the contract behavior is caused by the fact that transition guards with such variables are not evaluated; instead, both their values are assumed possible and state space exploration includes execution traces with and without each guarded transition. 
In essence, we analyze a more abstract model of the contract, with a set of reachable states and traces that is a superset of the set of states (respectively, traces) of the actual contract. 
As an example, let us consider the function in Figure~\ref{fig:overappr}.

\begin{figure}[h!]
\centering
\begin{lstlisting}[style=customjava] 
void fn(int x) {
   if (x < 0) {
       ... 	(1)
   }
   if (x > 0) {
       ... 	(2)
   }
}
\end{lstlisting}
 \caption{Code example for overapproximation.}
 \label{fig:overappr}
\end{figure}

 An overapproximation of the function's execution includes traces where both lines (1) and (2) are visited, even though they cannot both be satisfied by the same values of \text{x}. Note that abstraction is not necessary for variables that are independent of environment input (e.g. iteration counters of known range). These are updated in the model as they are calculated by contract statements.


We also apply abstraction to time variables (e.g. the \texttt{now} variable in the Blind Auction) using a slightly different approach. Although we need to know which transitions get invalidated as time increases, we do not represent the time spent in each state, as this time can be arbitrarily high.
Therefore, for a time-guarded transition in the model, say from a state $s_x$, one of the following applies:
\begin{compactitem}
\item if the guard is of type $t\leq t_{max}$, checking that a time variable does not exceed a threshold, a loop transition is added to $s_x$, with an action $t=t_{max}+1$ that invalidates the guard. A deadlock may be found in traces where this invalidating loop is executed (e.g., if no other transitions are offered in $s_x$).

\item if the guard is of type $t>t_{min}$, checking that a time variable exceeds a threshold, an action $t$=$t_{min}$+1 is added to the guarded transition. This sets the time to the earliest point that next state can be reached (e.g., useful for checking bounded liveness properties.)

\end{compactitem}

This overapproximation has the following implications.

\textbf{Safety properties:} \textit{Safety properties that are fulfilled in the abstract model are also guaranteed in the actual system.} Each safety property checks 
the non-reachability of a set of 
erroneous states. If these states are unreachable in the abstract model, they will be unreachable in the concrete model, which contains a subset of the abstract model's states. This property type is useful for checking vulnerabilities in currency withdrawal functions (e.g., the ``DAO attack'').

\textbf{Liveness properties:} \textit{Liveness properties that are violated in the abstract model are also violated in the actual system.} Each liveness property checks 
that a set of states are reachable. If they are found unreachable (i.e., liveness violation) in the abstract model, they will also be unreachable in the concrete model. This property type is useful for ``Denial-of-Service'' vulnerabilities (Appendix~\ref{app:vulnerabilities}).

\textbf{Deadlock freedom:} States without enabled outgoing transitions are identified as deadlock states. If no deadlock states are reachable in the abstract model, they will not be reachable in the actual system. 



\subsection{VeriSolid-to-BIP Mapping}

Since both VeriSolid and BIP model contract behavior as transition systems, the transformation is a simple mapping between the transitions, states, guards, and actions of VeriSolid to the transitions, states, guards, and actions of BIP (see Appendix~\ref{app:BIPmodeling} for background on BIP). Because this is an one-to-one mapping, we do not provide a proof.  
Our translation algorithm performs a single-pass syntax-directed parsing of the user's VeriSolid input and collects values that are appended to the attributes list of the templates. Specifically, the following values are collected:
\begin{itemize}[nosep,topsep=0pt]
\item variables $v \in V$, where $type(v)$ is the data type of $v$ and~$name(v)$ is the variable name (i.e., identifier);
\item states $s \in {S}$;
\item  transitions $t \in$ $T$, where  $t^{name}$ is the transition (and corresponding port) name, $t^{from}$ and $t^{to}$ are the outgoing and incoming states, $a_t$ and $g_t$ are invocations to functions that implement the associated actions and guards.
\end{itemize}

\begin{figure}[h!]
\begin{align*}
& \code{atom type Contract}() \\
\forall v \in V : ~ & \ind \code {data } type(v)~ name(v) \\
\forall t \in T: ~ & \ind \code{export port synPort} ~ t^{name}\code{()} \\
& \ind \code{places } s_0, \ldots, s_{|S| -1} \\
& \ind \code{initial to } s_0 \\
\forall t \in T: ~ & \ind \code{on } t^{name} \code{ from } t^{from} \code{ to } t^{to} \\
& \ind \ind \code{ provided (} g_t \code{) do } 
\{ a_t \} \\
& \code{end}
\ind \end{align*}
\caption{BIP code generation template.}
\label{fig:BIPtemplate}
\end{figure}

Figure~\ref{fig:BIPtemplate} shows the BIP code template. 
We use $\code{fixed-width}$ font
for the generated output, and $\lbl{italic}$ font for elements that are replaced with input.





\begin{table}[t]
  \caption{Analyzed properties and verification results for the case study models.}
  \centering
  \begin{tabular}{ p{2.2cm} p{6.8cm} p{1.3cm} p{1.1cm} }
  
    \textbf{Case Study} & \textbf{Properties} & \textbf{Type}  & \textbf{Result} \\ \hline  \hline
    \vspace{-.8cm}
    \multirow{1}{2cm}{
    \begin{tabular}{p{2cm}}
        BlindAuction (initial)\\
        states: 54
    \end{tabular}} & 
    \hspace{.1cm}\begin{tabular}{p{6cm}}
    (i) \texttt{bid} cannot happen after \texttt{close}: \\
    \hspace{.4cm}$\mathtt{AG}\bigl(
  \mathit{close} \rightarrow \mathtt{AG} \lnot \mathit{bid}
  \bigr)$ 
    \end{tabular} & Safety & Verified \\
  
 &  \hspace{.1cm}\begin{tabular}{p{6cm}} 
 (ii) \texttt{cancelABB} or \texttt{cancelRB} cannot happen after \texttt{finish}:  \\
 \hspace{.3cm} 
 $\mathtt{AG}\bigl(
 {\centering \mathit{finish} \rightarrow \mathtt{AG} \lnot \bigl( \mathit{cancelRB} \lor \mathit{cancelABB} \bigr) } \bigr)$ \end{tabular}
  & \vspace{-.05cm} Safety & \vspace{-.05cm}Verified \\
  
  &  \hspace{.1cm}\begin{tabular}{p{6cm}} 
  (iii) \texttt{withdraw} can happen only after \texttt{finish}: \\ \hspace{.4cm} $\mathtt{A}\bigl[\:
  \lnot \mathit{withdraw} \; \mathtt{W} \; \mathit{finish} 
  \:\bigr] $ \end{tabular} & Safety & Verified \\
  
  &  \hspace{.1cm}\begin{tabular}{p{6cm}} 
  (iv)  \texttt{finish} can happen only after \texttt{close}: \\ \hspace{.3cm} $\mathtt{A}\bigl[\:
  \lnot \mathit{finish} \; \mathtt{W} \; \mathit{close} 
  \:\bigr] $ \end{tabular} & Safety & Verified \\ \hline

    \vspace{-.8cm}
     \multirow{2}{2cm}{
     \begin{tabular}{p{2cm}}
     BlindAuction (augmented)\\
     states: 161
     \end{tabular}
     } 
     & \hspace{.1cm} \begin{tabular}{p{6cm}}
     (v) \texttt{23} cannot happen after \texttt{18}: \\
     \hspace{.4cm} $\mathtt{AG}\bigl(
  \mathit{18} \rightarrow \mathtt{AG} \lnot \mathit{23}
  \bigr)$ \end{tabular} & Safety & Verified \\
  
 	&  \hspace{.1cm} \begin{tabular}{p{6cm}} 
 	(vi) if \texttt{21} happens, \texttt{21} can happen only \mbox{after} \texttt{24}: \\
 	\hspace{.4cm}  $\mathtt{AG}\bigl(
  \mathit{21} \rightarrow 
  \mathtt{AX} \; \mathtt{A} \bigl[ \lnot \mathit{21}  \; \mathtt{W} \; \bigr( \mathit{24}\bigr) \;  \bigr]
  \bigr)$ \end{tabular} & Safety & Verified \\ \hline


   \vspace{-.6cm}\begin{tabular}{p{2cm}}
    DAO attack\\
    states: 9
    \end{tabular}
     &  \hspace{.1cm} \begin{tabular}{p{6cm}}
     if \texttt{call} happens, \texttt{call} can happen only \mbox{after} \texttt{subtract}: \\ 
     \hspace{.4cm}$\mathtt{AG}\bigl(
  \mathit{call} \rightarrow 
  \mathtt{AX} \; \mathtt{A} \bigl[ \lnot \mathit{call}  \; \mathtt{W} \; \mathit{subtract}  \bigr]
  \bigr)$ \end{tabular}
   & \vspace{-.6cm} Safety & \vspace{-.6cm} Verified \\  \hline
  
  \vspace{-.4cm}\begin{tabular}{p{2.2cm}}
     King of Ether 1\\
     states: 10
     \end{tabular}
     &  
     \hspace{.2cm}\begin{tabular}{p{6cm}}
    \texttt{7} will eventually happen after \texttt{4}: 
     \\
     
     \hspace{.4cm} $\mathtt{AG}\bigl(
  \mathit{4} \rightarrow 
  \mathtt{AF} \;\mathit{7} \;
  \bigr)$ \end{tabular}
  
  & \vspace{-.4cm} Liveness & \vspace{-.4cm}Violated \\  \hline
  
     \vspace{-.4cm}\begin{tabular}{p{2.2cm}}
     King of Ether 2\\
     states: 10
     \end{tabular}
     &  
       \hspace{.2cm}\begin{tabular}{p{6cm}} 
     \texttt{8} will eventually happen after \texttt{fallback}: \\
     
     \hspace{.4cm}  $\mathtt{AG}\bigl(
  \mathit{fallback} \rightarrow 
  \mathtt{AF} \; \mathit{8} \;
  \bigr)$  \end{tabular}
  
  & \vspace{-.4cm} Liveness & \vspace{-.4cm} Violated \\  \hline

  \end{tabular}
  \label{tab:results}
\end{table}

\subsection{Verification Results}
\label{sec:verificationResults}

Table~\ref{tab:results} summarizes the properties and verification results. For ease of presentation, when properties include statements, we replace statements with  the augmented-transition numbers that we have added to Figures~\ref{fig:completeBlind},~\ref{fig:king1}, and~\ref{fig:king2} in Appendices~\ref{app:completeBlindAuctionAugmented} and~\ref{sec:king}. The number of states represents the reachable state space as evaluated by nuXmv.

\subsubsection{Blind Auction}
We analyzed both the initial and augmented models of the Blind Auction contract. On the initial model, we checked four safety properties (see Properties (i)--(iv) in Table~\ref{tab:results}). On the augmented model, which allows for more fine-grained analysis, we  checked two additional safety properties. All properties were verified to hold. The models were found to be deadlock-free and their state space was 
evaluated to 
54 and 161 states, respectively. The augmented model and generated code can be found in Appendix~\ref{sec:completeBlindAuction}.

\subsubsection{The DAO Attack}


We modeled a simplified version of the DAO contract. Atzei et al.~\cite{atzei2017survey} discuss two different vulnerabilities exploited on DAO and present different attack scenarios. Our verified safety property (Table~\ref{tab:results}) excludes the possibility of both attacks. The augmented model can be found in Appendix~\ref{sec:dao}.



\subsubsection{King of the Ether Throne}
\label{sec:kingOfTheEtherVerification}

For checking Denial-of-Service vulnerabilities, we  created models of two versions of the King of the Ether contract~\cite{atzei2017survey}, which are provided in Appendix~\ref{sec:king}. On  ``King of Ether 1,'' we checked a liveness property stating that crowning (transition $7$) will happen at some time after the compensation calculation (transition $4$). The property is violated by the following counterexample: 
$\mathit{fallback} \rightarrow
\mathit{4} \rightarrow \mathit{5}$.
A second liveness property, which states that the crowning will happen at some time after fallback fails in ``King of Ether 2.'' A counterexample of the property violation is the following:
$\mathit{fallback} \rightarrow
\mathit{4}$. 
Note that usually many counterexamples may exist for the same  violation.

\subsubsection{Resource Allocation}
We have additionally verified a larger smart contract that acts as the core of a blockchain-based platform for transactive energy systems. The reachable state space, as evaluated by nuXmv, is $3,487$. Properties were verified or shown to be violated within seconds. Due to space limitations, we present the verification results in Appendix~\ref{app:transax}.

 \section{Related Work}
\label{sec:related}

\Aron{Make sure that we do not say anything negative about any PC member's paper!}
Here, we present a brief overview of related work.
We provide a more detailed discussion in Appendix~\ref{app:extended_related}.


Motivated by the large number of smart-contract vulnerabilities in practice, researchers have investigated and established taxonomies for common types of contract vulnerabilities~\cite{atzei2017survey,luu2016making}. 
To find vulnerabilities in existing contracts, 
both verification and vulnerability discovery are considered in the literature~\cite{parizi2018empirical}. 
In comparison, the main advantage of our model-based approach is that it allows developers to specify desired properties with respect to a high-level model instead of, e.g., EVM bytecode, and also provides verification results and counterexamples in a developer-friendly, easy to understand, high-level form.
Further, our approach allows verifying whether a contract satisfies all desired security properties instead of detecting certain types of vulnerabilities; hence, it can detect atypical vulnerabilities.

Hirai performs a formal verification of a smart contract used by the Ethereum Name Service~\cite{hirai2016formal} and defines the complete instruction set of the Ethereum Virtual Machine (EVM) in Lem, a language that can be compiled for interactive theorem provers, which enables proving certain safety properties for existing contracts~\cite{hirai2017defining}.
Bhargavan et al.\ outline a framework for verifying the safety and correctness of Ethereum contracts based on translating Solidity and EVM bytecode contracts into $F^{*}$~\cite{bhargavan2016short}.
Tsankov et al. introduce a security analyzer for Ethereum contracts, called \textsc{Securify}, which symbolically encodes the dependence graph of a contract in stratified Datalog~\cite{jeffrey1989principles} and then uses off-the-shelf solvers to check the satisfaction of properties~\cite{tsankov2018securify}.
Atzei et al.\ prove the well-formedness properties of the Bitcoin blockchain have also been proven using a formal model~\cite{atzei2018formal}.
Techniques from runtime verification are used to detect and recover from violations at runtime~\cite{ellul2018runtime,colombo2018contracts}.

Luu et al.\ provide a tool called \textsc{Oyente}, which can analyze  contracts and detect certain typical security vulnerabilities~\cite{luu2016making}.
Building on \textsc{Oyente}, Albert et al. introduce the \textsc{EthIR} framework, which can produce a rule-based representation of bytecode, enabling the application of existing analysis to infer properties of the EVM codee~\cite{albert2018ethir}.
Nikolic et al. present the \textsc{MAIAN} tool for detecting three types of vulnerable contracts, called prodigal, suicidal and greedy~\cite{nikolic2018finding}.
Fr{\"o}wis and B{\"o}hme define a heuristic indicator of control flow immutability to quantify the prevalence of contractual loopholes based on modifying the control flow of Ethereum contracts~\cite{frowis2017code}.
%
Brent et al. introduce a security analysis framework for
Ethereum contracts, called \textsc{Vandal},
which converts EVM bytecode to semantic relations, which are then analyzed to detect vulnerabilities described in the Souffl\'e language~\cite{brent2018vandal}.
Mueller presents \textsc{Mythril}, a security analysis tool for Ethereum smart contracts with a symbolic execution backend~\cite{mueller2018smashing}.
Stortz introduces \textsc{Rattle}, a static analysis framework for EVM bytecode~\cite{stortz2018rattle}.

%
%

Researchers also focus on providing formal operational semantics for EVM bytecode and Solidity language~\cite{hildenbrandt2017kevm,grishchenko2018semantic,grishchenko2018semantic2,yang2018lolisa,jiao2018executable}.
Common design patterns in Ethereum smart contracts are also identified and studied by multiple research efforts~\cite{bartoletti2017empirical,wohrer2018design}.
Finally, to facilitate development, researchers have also introduced a functional smart-contract language~\cite{oconnor2017simplicity}, an approach for semi-automated translation of human-readable contract representations into computational
equivalents~\cite{frantz2016institutions}, a logic-based smart-contract model~\cite{hu2018method}.

 \section{Conclusion}
\label{sec:concl}

%
We presented an end-to-end framework that allows the generation of correct-by-design contracts by performing 
a set of equivalent transformations. First, we generate an augmented transition system from an initial transition system, based on the operational semantics of supported Solidity statements (Appendix~\ref{app:soliditySemantics}). We have proven that the two transition systems are observationally equivalent (Section~\ref{sec:proof}). Second, we generate the BIP transition system from the augmented transition system through a direct one-to-one mapping. Third, we generate the NuSMV  transition system from the BIP system (shown to be observationally equivalent in~\cite{noureddine2014reduction}). Finally, we generate functionally equivalent Solidity code, based on the operational semantics of the transition system (Appendix~\ref{app:transitionSemantics}).

To the best of our knowledge, VeriSolid is the first framework to promote a model-based, correctness-by-design approach for blockchain-based smart contracts. Properties established at any step of the VeriSolid design flow are preserved in the resulting smart contracts, guaranteeing their correctness. 
VeriSolid fully automates the process of verification and code generation,  while enhancing usability by providing easy-to-use graphical editors for the specification of transition systems and natural-like language templates for the specification of formal properties.
By performing verification early at design time, we  provide a cost-effective approach; fixing bugs later in the development process can be very expensive. 
%
%
%
Our verification approach can detect typical vulnerabilities, but it may also detect any violation of required properties. Since our tool applies verification at a high-level, it can provide meaningful feedback to the developer when a property is not satisfied, which would be much harder to do at bytecode level.
%
%
Future work includes extending the approach to model and generate correct-by-design \emph{systems of interacting smart contracts}.

\clearpage
\appendix

\section{Formalisms}

\subsection{Supported Solidity Subset}
\label{app:soliditySubset}

Here, we define the subset of Solidity that VeriSolid supports. First, let us introduce the following notation:

\begin{itemize}
    \item Let $\mathbb{T}$ denote the set of Solidity types;
    \item let $\mathbb{I}$ denote the set of valid Solidity identifiers;
    \item let $\mathbb{D}$ denote the set of Solidity event and custom type definitions;
    \item let $\mathbb{E}$ denote the set of Solidity expressions;
    \item let $\mathbb{C}$ denote the set of Solidity expressions without side effects (i.e., expression whose evaluation does not change storage, memory, balances, etc.);
    \item let $\mathbb{S}$ denote the set of supported Solidity statements.
\end{itemize}

We define the set of supported event ($\langle event \rangle$) and custom type ($\langle struct \rangle$) definitions $\mathbb{D}$ as follows:
{\small
\begin{align*}
\langle event \rangle ::= \code{event} ~ @identifier ~ \code{(}& ~ \big(@type ~ @identifier \\
 & (\code{,} ~ @type ~ @identifier)*\big)? ~ \code{);} 
\end{align*}
}
{\small
\begin{align*}
\langle struct \rangle ::= \code{struct} ~ @identifier ~ \code{\{} ~ (@type ~ @identifier ~ \code{;})* ~ \code{\}} 
\end{align*}
}

We let $\mathbb{E}$ denote the set of Solidity expressions.
We let $\mathbb{C}$ denote the following subset of Solidity expressions, which do not have any side effects:
{\small
\begin{align*}
\langle pure \rangle ::=
  &| ~ \langle variable \rangle \\
  &| ~ @{constant} \\
  &| ~ \code{(} ~ \langle pure \rangle ~ \code{)} \\
  &| ~ \langle unary \rangle ~ \langle pure \rangle \\
  &| ~ \langle pure \rangle ~ \langle operator \rangle ~ \langle pure \rangle 
\end{align*}
\vspace{-1.5em}
\begin{align*}
\langle variable \rangle ::=
  &| ~ @{identifier} \\
  &| ~ \langle variable \rangle ~ \code{.} ~ @{identifier} \\
  &| ~ \langle variable \rangle ~  \code{[} ~ \langle pure \rangle ~ \code{]}
\end{align*}
\vspace{-1.5em}
\begin{align*}
\langle operator \rangle ::= ~ \code{==} ~ | ~ \code{!=} ~ | ~ \code{<} ~ | ~ \code{>} ~ | ~ \code{>=} ~ | ~ \code{<=} \\
~ | ~ \code{+} ~ | ~ \code{*} ~ | ~ \code{-} ~ | ~ \code{/} ~ | ~ \code{\%} ~ | ~ \code{\&\&} ~ | ~ \code{||} 
\end{align*}
\vspace{-1.5em}
\begin{align*}
\langle unary \rangle ::= ~ \code{!} ~ | ~ \code{+} ~ | ~ \code{-}
\end{align*}
}

VeriSolid supports the following types of statements:
\begin{compactitem}
\item variable declarations (e.g., \texttt{\small int32 value = 0;} and \texttt{\small address from = msg.sender;}),
\item expressions (e.g., \texttt{\small amount = balance[msg.sender];} \\ or \texttt{\small msg.sender.transfer(amount);}),
\item event statements (e.g., \texttt{\small emit Deposit(amount, msg.sender);}),
\item \texttt{\small return} statements (e.g., \texttt{\small return;} and \texttt{\small return amount;}),
\item \texttt{\small if} and \texttt{\small if} ... \texttt{\small else} selection statements (including \texttt{\small if} ... \texttt{\small else if} ... and so on),
\item \texttt{\small for} and \texttt{\small while} loop statements, 
\item compound statements (i.e., \texttt{\small \{ \textit{statement1} \textit{statement2} ... \}}).
\end{compactitem}
We define the formal grammar of the subset of supported Solidity statements $\mathbb{S}$ as follows:
{\small
\begin{align*}
\langle statement& \rangle ::= \\
&| ~ \langle declaration \rangle ~ \code{;}  \\
&| ~ @expression ~ \code{;} \\
&| ~ \code{emit } @identifier \code{(} ~ \big(@expression \\ & \hspace{2em} (\code{,} ~ @expression)* \big)? \code{);} \\
&| ~ \code{return} ~ (@pure)? ~ \code{;} \\
&| ~ \code{if (} ~ @expression ~ \code{)} ~ \langle statement \rangle \\ & \hspace{2em} (\code{else} ~ \langle statement \rangle)? \\
&| ~ \code{for (} ~ \langle declaration \rangle ~ \code{;} ~ @expression ~ \code{;} \\ & \hspace{2em} @expression ~ \code{)} ~ \langle statement \rangle \\
&| ~ \code{while (} ~ @expression \code{)} ~ \langle statement \rangle \\
&| ~ \code{\{} ~ ( \langle statement \rangle )* ~ \code{\}}
\end{align*}
\begin{align*}
\langle declaration \rangle ::= ~ @type ~ @{identifier} ~ (\code{=} ~ @expression)? 
\end{align*}
}
\noindent where $@expression \in \mathbb{E}$ is a primary Solidity expression, which may include function calls, transfers, etc.,
while $@pure \in \mathbb{C}$ is a Solidity expression without side effects.

\subsection{Operational Semantics of the Transition System}
\label{app:transitionSemantics}

We let $\Psi$ denote the state of the ledger, which includes account balances, values of state variables in all contracts, number and timestamp of the last block, etc.
During the execution of a transition, the execution state $\sigma = \{\Psi, M\}$ also includes the memory and stack state~$M$. 
To handle return statements and exceptions, we also introduce an execution status, which is equal to $E$ when an exception has been raised, $R[v]$ when a return statement has been executed with value $v$ (i.e., \texttt{return} $v$), and $N$ otherwise.
Finally, we let $\Eval{\sigma}{\Exp} \rightarrow \langle (\hat{\sigma}, x), v\rangle$ signify that the evaluation of a Solidity expression $\Exp$ in execution state~$\sigma$ yields value $v$ and---as a side effect---changes the execution state to~$\hat{\sigma}$ and the execution status to $x$.

A transition is triggered by providing a transition (i.e., function) $\textit{name} \in \mathbb{I}$ and a list of parameter values $v_1, v_2, \ldots$.
The normal execution of a transition without returning any value, which takes the ledger from state $\Psi$ to  $\Psi'$ and the contract from state $s \in S$ to  $s' \in S$, is captured by the \textsc{TRANSITION} rule:

\OSRule{
t \in T , \hspace{0.8em}  
\textit{name} = t^{name}  , \hspace{0.8em}
s = t^{\textit{from}}  \\
M = \Params{t}{v_1, v_2, \ldots} , \hspace{0.8em}
\sigma = (\Psi, M) \\
\Eval{\sigma}{g_t} \rightarrow \langle (\hat{\sigma}, N), \True \rangle \\
\langle (\hat{\sigma}, N), a_t \rangle \rightarrow \langle (\hat{\sigma}', N), \cdot \rangle \\
\hat{\sigma}' = (\Psi', M') , \hspace{0.8em}
s' = t^{\textit{to}}
}{
(\Psi, s), \textit{name}\,(v_1, v_2, \ldots)
}{
(\Psi', s', \cdot) 
}{
TRANSITION
}

This rule is applied if there exists a transition $t$ whose name $t^{\textit{name}}$ is $\textit{name}$ and whose source state $t^{\textit{from}}$ is the current contract state $s$ (first line).
The execution state $\sigma$ is initialized by taking the parameter values $\Params{t}{v_1, v_2, \ldots}$ and the current ledger state $\Psi$ (second line).
If the guard condition $g_t$ evaluates  $\Eval{\sigma}{g_t}$ in the current state $\sigma$ to $\True$ without any exceptions (third line), then the action statement $a_t$ of the transition is executed (fourth line), which results in an updated execution state $\hat{\sigma}'$ (see statement rules in Appendix~\ref{app:soliditySemantics}).
Finally, if the execution status resulting from the action is normal $N$ (i.e., no exception was thrown), then the updated ledger state $\Psi'$ and updated contract state $s'$ (fifth line) are made permanent.

The normal execution of a transition that returns a value is captured by the \textsc{TRANSITION-RET} rule:

\OSRule{
t \in T , \hspace{0.8em}  
\textit{name} = t^{name}  , \hspace{0.8em}
s = t^{\textit{from}}  \\
M = \Params{t}{v_1, v_2, \ldots} , \hspace{0.8em}
\sigma = (\Psi, M) \\
\Eval{\sigma}{g_t} \rightarrow \langle (\hat{\sigma}, N), \True \rangle \\
\langle (\hat{\sigma}, N), a_t \rangle \rightarrow \langle (\hat{\sigma}', R[v]), \cdot \rangle  \\
\hat{\sigma}' = (\Psi', M') , \hspace{0.8em}
s' = t^{\textit{to}}
}{
(\Psi, s), \textit{name}\,(v_1, v_2, \ldots)
}{
(\Psi', s', v) 
}{
TRANSITION-RET
}

This rule is applied if the transition action $a_t$ finishes with a \texttt{return} $v$ statement, resulting in execution status $R[v]$. 

If the transition $t$ by name $t^{\textit{name}} = \textit{name}$ exists, but its source state $t^{\textit{from}}$ is not $s$, then the transition is not executed, which is captured by the \textsc{TRANSITION-WRO} rule:

\OSRule{
t \in T , \hspace{0.8em}  
\textit{name} = t^{name}  , \hspace{0.8em}
s \neq t^{\textit{from}}  
}{
(\Psi, s), \textit{name}\,(v_1, v_2, \ldots)
}{
(\Psi, s, \cdot) 
}{
TRANSITION-WRO
}

Similarly, if the guard condition $g_t$ of the transition evaluates $\Eval{\sigma}{g_t}$ to $\False$, then the transition is reverted, , which is captured by the \textsc{TRANSITION-GRD} rule:

\OSRule{
t \in T , \hspace{0.8em}  
\textit{name} = t^{name}  , \hspace{0.8em}
s = t^{\textit{from}}  \\
M = \Params{t}{v_1, v_2, \ldots} , \hspace{0.8em}
\sigma = (\Psi, M) \\
\Eval{\sigma}{g_t} \rightarrow \langle (\hat{\sigma}, N), \False \rangle 
}{
(\Psi, s), \textit{name}\,(v_1, v_2, \ldots)
}{
(\Psi, s, \cdot) 
}{
TRANSITION-GRD
}

If an exception is raised during the evaluation $\Eval{\sigma}{g_t}$ of the guard condition $g_t$ (i.e., if the execution status becomes $E$), then the transition is reverted, which is captured by the \textsc{TRANSITION-EXC1} rule:

\OSRule{
t \in T , \hspace{0.8em}  
\textit{name} = t^{name}  , \hspace{0.8em}
s = t^{\textit{from}}  \\
M = \Params{t}{v_1, v_2, \ldots} , \hspace{0.8em}
\sigma = (\Psi, M) \\
\Eval{\sigma}{g_t} \rightarrow \langle (\hat{\sigma}, E), x \rangle 
}{
(\Psi, s), \textit{name}\,(v_1, v_2, \ldots)
}{
(\Psi, s, \cdot)  
}{
TRANSITION-EXC1
}

Similarly, if an exception is raised during the execution of the transition action $a_t$, then the transition is reverted, which is captured by the \textsc{TRANSITION-EXC2} rule:

\OSRule{
t \in T , \hspace{0.8em}  
\textit{name} = t^{name}  , \hspace{0.8em}
s = t^{\textit{from}}  \\
M = \Params{t}{v_1, v_2, \ldots} , \hspace{0.8em}
\sigma = (\Psi, M) \\
\Eval{\sigma}{g_t} \rightarrow \langle (\hat{\sigma}, N), \True \rangle \\
\langle (\hat{\sigma}, N), a_t \rangle \rightarrow \langle (\hat{\sigma}', E), \cdot \rangle \\
}{
(\Psi, s), \textit{name}\,(v_1, v_2, \ldots)
}{
(\Psi, s, \cdot)  
}{
TRANSITION-EXC2
}

On the other hand, if there exists no transition by the name \textit{name}, then the fallback action $a_F$ is executed, which is captured by the \textsc{TRANSITION-FAL} rule:

\OSRule{
\forall t \in T: \textit{name} \neq t^{name}  \\
\sigma = (\Psi, \emptyset) \\
\langle (\hat{\sigma}, N), a_F \rangle \rightarrow \langle (\hat{\sigma}', x), \cdot \rangle, \hspace{0.8em} x \neq E \\
\hat{\sigma}' = (\Psi', y) 
}{
(\Psi, s), \textit{name}\,(v_1, v_2, \ldots)
}{
(\Psi', s, \cdot)  
}{
TRANSITION-FAL
}

Finally, if an exception is raised during the execution of the fallback action $a_F$, then the transition is reverted, which is captured by the \textsc{TRANSITION-EXC3} rule:

\OSRule{
\forall t \in T: \textit{name} \neq t^{name}  \\
\sigma = (\Psi, \emptyset) \\
\langle (\hat{\sigma}, N), a_F \rangle \rightarrow \langle (\hat{\sigma}', E), \cdot \rangle
}{
(\Psi, s), \textit{name}\,(v_1, v_2, \ldots)
}{
(\Psi, s, \cdot)
}{
TRANSITION-EXC3
}

\subsection{Operational Semantics of Supported Solidity Statements}
\label{app:soliditySemantics}
We build on the  small-step operational semantics for Solidity defined in~\cite{jiao2018executable}, which enables us to reason about one computational step at a time. We have extended the semantics of~\cite{jiao2018executable} to support exceptions and return values.

We present the semantics of each supported Solidity statement as one or more rules.
Each rule takes an execution state $\sigma$, an execution status $\in \{N, E, R[v]\}$, and a statement $\Stmt \in \mathbb{S}$, and maps them to a new execution state, a new execution status, and a statement that remains to be executed (or $\cdot$ if no statements are left to be executed). 

We start with basic rules that apply to every statement.
If an exception has been raised or if a \texttt{return} statement has been executed, then no further statements should be executed, which is captured by the \textsc{SKIP-EXC} and \textsc{SKIP-RET} rules:

\OSRule{
}{
(\sigma, E), \Stmt
}{
(\sigma, E), \cdot
}{
SKIP-EXC
}

\OSRule{
}{
(\sigma, R[v]), \Stmt
}{
(\sigma, R[v]), \cdot
}{
SKIP-RET
}

A \texttt{return} statement changes the execution status to $R[\cdot]$, skipping all remaining statements, which is captured by the \textsc{RETURN} rule:

\OSRule{
}{
(\sigma, N), \texttt{return;} 
}{
(\sigma, R[\cdot]), \cdot 
}{
RETURN
}

To return a value $v$, a \texttt{return} $\Exp$ statement changes the execution status to $R[v]$, which is captured by the \textsc{RETURN-VAL} rule:

\OSRule{
\Eval{\sigma}{\Exp} \rightarrow \langle{(\sigma', N), v}\rangle
}{
(\sigma, N), \texttt{return \Exp;} 
}{
(\sigma', R[v]), \cdot 
}{
RETURN-VAL
}

If an exception is raised during the evaluation of $\Eval{\sigma}{\Exp}$, then execution status is changed to $E$, which is captured by the \textsc{RETURN-EXC} rule:

\OSRule{
\Eval{\sigma}{\Exp} \rightarrow \langle{(\sigma', E), v}\rangle
}{
(\sigma, N), \texttt{return \Exp;} 
}{
(\sigma', E), \cdot 
}{
RETURN-EXC
}

A compound statement (i.e., a list of statements enclosed in braces $\{$ and $\}$) is executed by executing inner statements one after another, which is captured by the \textsc{COMPOUND} rule:

\OSRule{
\langle (\sigma, N), \Stmt_1 \rangle \rightarrow \langle (\sigma_1, x_1), \cdot \rangle \\
\langle (\sigma_1, x_1), \Stmt_2 \rangle \rightarrow \langle (\sigma_2, x_2), \cdot \rangle \\
\ldots \\
\langle (\sigma_{n-1}, x_{n-1}), \Stmt_n \rangle \rightarrow \langle (\sigma', x), \cdot \rangle 
}{
(\sigma, N), \texttt{\{} \Stmt_1 ~ \Stmt_2 ~ \ldots ~ \Stmt_n \texttt{\}} 
}{
(\sigma', x), \cdot
}{
COMPOUND
}

\subsubsection{Loop Statements}

A \texttt{while} loop statement evaluates its condition $\Exp$ and if its $\False$, skips the execution of the body statement $\Stmt$, which is captured by the \textsc{WHILE1} rule:

\OSRule{
\Eval{\sigma}{\Exp} \rightarrow \langle (\sigma', N), \False \rangle
}{
(\sigma, N), ~ \texttt{while(}\Exp\texttt{)} \, \Stmt
}{
(\sigma', N), \cdot
}{
WHILE1
}

Similarly, if the evaluation of the loop condition $\Exp$ results is an exception, then execution of the body statement $\Stmt$ is skipped, which is captured by the \textsc{WHILE-EXC} rule:

\OSRule{
\Eval{\sigma}{\Exp} \rightarrow \langle (\sigma', E), x \rangle
}{
(\sigma, N), ~ \texttt{while(}\Exp\texttt{)} \, \Stmt
}{
(\sigma', E), \cdot
}{
WHILE-EXC
}

On the other hand, if the loop condition $\Exp$ is $\True$, then the body statement $\Stmt$ is executed, which is captured by the \textsc{WHILE2} rule:

\OSRule{
\Eval{\sigma}{\Exp} \rightarrow \langle (\hat{\sigma}, N), \True \rangle
\\
\langle (\hat{\sigma}, N), \Stmt \rangle \rightarrow \langle (\hat{\sigma}', x), \cdot \rangle
}{
(\sigma, N), ~ \texttt{while(}\Exp\texttt{)} \, \Stmt
}{
(\hat{\sigma}', x),  ~ \texttt{while(}\Exp\texttt{)} \, \Stmt
}{
WHILE2
}

A \texttt{for} loop statement can be reduced to a \texttt{while} loop, which is captured by the \textsc{FOR} rule:

\OSRuleBreak{
\langle (\sigma, N), \Stmt_I \rangle \rightarrow \langle (\sigma', x), \cdot \rangle
}{
(\sigma, N), ~ \texttt{for(}\Stmt_I\texttt{;}\, \Exp_C \texttt{;}\, \Stmt_A \texttt{)} ~ \Stmt_B
}{
(\sigma', x),  ~ \texttt{while(}\Exp_C\texttt{) \{} \, \Stmt_B ~ \Stmt_A \texttt{\}}
}{
FOR
}

\subsubsection{Selection Statements}

An \texttt{if} statement is captured by the \textsc{IF1}, \textsc{IF2}, and \textsc{IF-EXC} rules:

\OSRule{
\Eval{\sigma}{\Exp} \rightarrow \langle (\hat{\sigma}, N), \True \rangle
\\
\langle (\hat{\sigma}, N), \Stmt \rangle \rightarrow \langle (\hat{\sigma}', x), \cdot \rangle
}{
(\sigma, N), ~ \texttt{if(}\Exp\texttt{)} \, \Stmt
}{
(\hat{\sigma}', x),  \cdot
}{
IF1
}

\OSRule{
\Eval{\sigma}{\Exp} \rightarrow \langle (\hat{\sigma}, N), \False \rangle
}{
(\sigma, N), ~ \texttt{if(}\Exp\texttt{)} \, \Stmt
}{
(\hat{\sigma}, N),  \cdot
}{
IF2
}

\OSRule{
\Eval{\sigma}{\Exp} \rightarrow \langle (\hat{\sigma}, E), x \rangle
}{
(\sigma, N), ~ \texttt{if(}\Exp\texttt{)} \, \Stmt
}{
(\sigma, E),  \cdot
}{
IF-EXC
}

Similarly, an \texttt{if} $\ldots$ \texttt{else} statement is captured by three rules, \textsc{IFELSE1}, \textsc{IFELSE2}, and \textsc{IFELSE-EXC}:

\OSRule{
\Eval{\sigma}{\Exp} \rightarrow \langle (\hat{\sigma}, N), \True \rangle
\\
\langle (\hat{\sigma}, N), \Stmt_1 \rangle \rightarrow \langle (\hat{\sigma}', x), \cdot \rangle
}{
(\sigma, N), ~ \texttt{if(}\Exp\texttt{)} \, \Stmt_1 ~ \texttt{else} ~ \Stmt_2
}{
(\hat{\sigma}', x),  \cdot
}{
IFELSE1
}

\OSRule{
\Eval{\sigma}{\Exp} \rightarrow \langle (\hat{\sigma}, N), \False \rangle \\
\langle (\hat{\sigma}, N), \Stmt_2 \rangle \rightarrow \langle (\hat{\sigma}', x), \cdot \rangle
}{
(\sigma, N), ~ \texttt{if(}\Exp\texttt{)} \, \Stmt_1 ~ \texttt{else} ~ \Stmt_2
}{
(\hat{\sigma}', x),  \cdot
}{
IFELSE2
}

\OSRule{
\Eval{\sigma}{\Exp} \rightarrow \langle (\hat{\sigma}, E), x \rangle
}{
(\sigma, N), ~ \texttt{if(}\Exp\texttt{)} \, \Stmt_1 ~ \texttt{else} ~ \Stmt_2
}{
(\sigma, E),  \cdot
}{
IFELSE-EXC
}

\subsubsection{Miscellaneous Statements}

An expression statement is captured by the \textsc{EXPRESSION} rule:

\OSRule{
\Eval{\sigma}{\Exp} \rightarrow \langle{(\sigma', x), v}\rangle
}{
(\sigma, N), \texttt{\Exp;} 
}{
(\sigma', x), \cdot 
}{
EXPRESSION
}

A variable declaration statement is captured by the \textsc{VARIABLE} or \textsc{VARIABLE-ASG} rule:

\OSRule{
\Decl{\sigma}{\Type, \textnormal{Name}} \rightarrow \langle{(\sigma', x)}\rangle
}{
(\sigma, N), \texttt{\Type \,\,\textnormal{Name};} 
}{
(\sigma', x), \cdot 
}{
VARIABLE
}

\OSRuleBreak{
\Eval{\sigma}{\Exp} \rightarrow \langle(\sigma', x), v\rangle 
}{
(\sigma, N), \texttt{\Type ~\textnormal{Name} = \Exp;} 
}{
(\sigma', x), \texttt{\{ \Type\,\,\textnormal{Name}; \textnormal{Name}  =  $v$; \}} 
}{
VARIABLE-ASG
}

\noindent 
where $\Decl{\sigma}{\Type, \textnormal{Name}}$ introduces a variable into the namespace (and extends memory when necessary for memory-type variables).

An event statement is captured by the \textsc{EVENT} and \textsc{EVENT-EXC} rules:

\OSRule{
\Eval{\sigma}{\Exp_1} \rightarrow \langle{(\sigma_1, N), v_1}\rangle \\
\ldots \\
\Eval{\sigma_{n-1}}{\Exp_n} \rightarrow \langle{(\sigma_n, N), v_n}\rangle \\
\Log(\sigma_n, (name, v_1, \ldots, v_n)) \rightarrow  (\sigma', N) \\
}{
(\sigma, N), ~\texttt{emit}~\textit{name}\,(\Exp_1, \ldots, \Exp_n)\texttt{;} 
}{
(\sigma', x), \cdot 
}{
EVENT
}

\OSRule{
\Eval{\sigma}{\Exp_1} \rightarrow \langle{(\sigma_1, x_1), v_1}\rangle \\
\ldots \\
\Eval{\sigma_{n-1}}{\Exp_n} \rightarrow \langle{(\sigma_n, x_n), v_n}\rangle \\
\Log(\sigma_n, (name, v_1, \ldots, v_n)) \rightarrow  (\sigma', y) \\
x_1 = E \vee \ldots \vee x_n = E \vee y = E
}{
(\sigma, N), ~\texttt{emit}~\textit{name}\,(\Exp_1, \ldots, \Exp_n)\texttt{;} 
}{
(\sigma', x), \cdot 
}{
EVENT-EXC
}

\noindent
where $\Log$ records the specified values on the blockchain.

\section{Templates and CTL for Property Specification}
\label{app:templates}


\subsection{Background on CTL }
\label{app:ctl}

For the specification of properties, we use Computation Tree Logic (CTL). We only provide
a brief overview, referring the reader to the classic textbook
\cite{BaierKatoen2008} for a complete and formal presentation.
CTL formulas specify properties of execution trees generated by transitions systems.
The formulas are built from atomic predicates that represent 
transitions and statements of the
transition system, using several operators, such as $\mathtt{EX}$, $\mathtt{AX}$,
$\mathtt{EF}$, $\mathtt{AF}$, $\mathtt{EG}$, $\mathtt{AG}$ (unary) and
$\mathtt{E}[\cdot\,\mathtt{U}\,\cdot]$,
$\mathtt{A}[\cdot\,\mathtt{U}\,\cdot]$,
$\mathtt{E}[\cdot\,\mathtt{W}\,\cdot]$,
$\mathtt{A}[\cdot\,\mathtt{W}\,\cdot]$ (binary).  Each operator
consists of a quantifier on the branches of the tree and a temporal
modality, which together define when in the execution the operand
sub-formulas must hold.  The intuition behind the letters is the
following: the branch quantifiers are $\mathtt{A}$ (for ``All'') and
$\mathtt{E}$ (for ``Exists''); the temporal modalities are
$\mathtt{X}$ (for ``neXt''), $\mathtt{F}$ (for ``some time in the
Future''), $\mathtt{G}$ (for ``Globally''), 
$\mathtt{U}$ (for ``Until'') and $\mathtt{W}$ (for ``Weak until'').  
A property is satisfied if it holds in the initial state of the transition systems.  
For instance, the formula
$\mathtt{A}[p\,\mathtt{W}\,q]$ specifies that in \emph{all execution
  branches} the predicate $p$ must hold \emph{up to the first state}
(not including this latter), where the predicate $q$ holds.  
Since we used the weak until operator $\mathtt{W}$, if $q$ never
holds, $p$ must hold forever.
As soon
as $q$ holds in one state of an execution branch, $p$ does not need to hold
anymore, even if $q$ does not hold.  On the contrary, the formula
$\mathtt{AG}\,\mathtt{A}[p\,\mathtt{W}\,q]$ specifies that the
subformula $\mathtt{A}[p\,\mathtt{W}\,q]$ must hold in \emph{all
  branches at all times}.  Thus, $p$ must hold whenever $q$ does not hold, i.e., $\mathtt{AG}\,\mathtt{A}[p\,\mathtt{W}\,q] = \mathtt{AG}\,(p \lor q)$.
  
\subsection{Templates and Corresponding CTL formulas}
Tables~\ref{tab:safety} and~\ref{tab:liveness} contain the full list of our natural language-like templates and their corresponding CTL formulas. We use $p$, $q$, and $r$ for simplicity, to denote the transition and statement sets, i.e., \textbf{$\langle{\textit{Transitions}\,\cup\,\textit{Statements}\rangle}$}. 

\begin{table}
\centering
\caption{Safety property templates}
\label{tab:safety}
\begin{tabular}{|c|c|}
\hline
Template & \multicolumn{1}{c|}{CTL formula} \\ 
\hline
\textbf{p} can never happen after \textbf{q} & $\mathtt{AG}(\textbf{q}$ $\rightarrow$ $\mathtt{AG}$ $(\neg \textbf{p}))$ \\
\rowcolor{TableRowGray} \textbf{p} can happen only after \textbf{q} & $\mathtt{A}[ \neg$ $\textbf{p}$ $\mathtt{W}$ $\textbf{q}]$ \\
if \textbf{p} happens, \textbf{q} can happen only after \textbf{r} happens & $\mathtt{AG}(\textbf{p}$ $\rightarrow$ $\mathtt{AX}$ $\mathtt{A}$ $[\neg$ $\textbf{q}$ $\mathtt{W}$ $\textbf{r}])$ \\
\rowcolor{TableRowGray} \textbf{p} can never happen & $\mathtt{AG}($ $ \neg \textbf{p})$ \\
\textbf{p} can never happen before \textbf{q} & $\mathtt{A}[ \neg$ $\textbf{p}$ $|$ $\mathtt{AG}($ $ \neg \textbf{q})$ $\mathtt{W}$ $\textbf{q}]$ \\
\hline
\end{tabular}
\end{table}

\begin{table}
\centering
\caption{Liveness property templates}
\label{tab:liveness}
\begin{tabular}{|c|c|}
\hline
Template & \multicolumn{1}{c|}{CTL formula} \\ 
\hline
\textbf{p} will eventually happen after \textbf{q} & $\mathtt{AG}$ $(\textbf{q}$ $\rightarrow$ $\mathtt{AF}$ $(\textbf{p}))$ \\
\rowcolor{TableRowGray} \textbf{p} will eventually happen & $\mathtt{AF}($ $ \textbf{p})$ \\
\hline
\end{tabular}
\end{table}

\section{Background}
\label{sec:background}

\subsection{Solidity Function Calls}
 \label{app:functionCalls}
Nested function calls in Solidity are the reason behind several identified vulnerabilities. 
We briefly describe how a smart contract can call a function of another contract or delegate execution. More information can be found in the Solidity documentation~\cite{solidityFunctionCalls}.
%
Firstly, a contract can call functions defined in another contract: 
\begin{itemize}[nosep,topsep=0pt]
\item \textit{addressOfContract}\texttt{.call(}\textit{data}\texttt{)}: Low-level call, for which the name and arguments of the invoked function must be specified in \textit{data} according to the Ethereum ABI.
The \texttt{call} method returns Boolean \texttt{true} if the execution was successful (or if there is no contract at the specified address) and \texttt{false} if it failed (e.g., if the invoked function threw an exception). 
\item \textit{contract}\texttt{.}\textit{function}\texttt{(}\textit{arg1}\texttt{,} \textit{arg2}\texttt{,} \textit{...}\texttt{)}: High-level call\footnote{Note that \textit{contract} is a reference to a Solidity contract that is available at compile time, while \textit{addressOfContract} is just a 160-bit address value.}, which may return a value as output on success. 
If the invoked method fails (or does not exist), an exception is raised in the caller, which means that all changes made by the caller are reverted, and the exception is automatically propagated up in the call hierarchy.
\end{itemize}

If the function specified for \texttt{call} does not exist, then the \emph{fallback} function of the callee is invoked. 
The fallback function does not have a name\footnote{For ease of presentation, we will refer to the fallback function using the name ``fallback'' in our models.} and arguments, and it cannot return anything.
A contract can have \emph{at most one} fallback function, and no function is executed if a fallback is not found (note that this does not constitute a failure).
The fallback function is also invoked if ether\footnote{Ether is the cryptocurrency provided by the Ethereum blockchain.} is sent to the contract using one of the two methods:
\begin{itemize}[nosep,topsep=0pt]
\item \textit{addressOfContract}\texttt{.send(}\textit{amount}\texttt{)}: Sends the specified amount of currency to the contract, invoking its fallback function (if there exists one).
If \texttt{send} fails (e.g., if the fallback function throws an exception), then it returns Boolean \texttt{false}; otherwise, it returns \texttt{true}.
\item \textit{addressOfContract}\texttt{.transfer(}\textit{amount}\texttt{)}: Similar to \texttt{send}, but raises an exception on failure, which is handled similar to a high-level function call failure.
\end{itemize}

Finally, a contract can also ``delegate'' execution to another contract using \textit{addressOfContract}\texttt{.delegatecall(}\textit{data}\texttt{)}.
Delegation is similar to a low-level \texttt{call}, but there is a fundamental difference: in this case, the function specified by \textit{data} is executed in the context of the caller (e.g., the function will see the contract variables of the caller, not the callee).
In other words, contracts may ``borrow code'' from other contracts using \texttt{delegatecall}, which enables the creation of libraries.

\subsection{Examples of Common Solidity Vulnerabilities}
\label{app:vulnerabilities}

Here, we discuss three examples of common types of vulnerabilities in Solidity smart contracts~\cite{sergey2017concurrent,atzei2017survey}.

\subsubsection{Re-Entrancy}
When a contract calls a function in another contract, the caller is blocked until the call returns. 
This allows the callee, who may be malicious, to take advantage of the intermediate state in which the caller is, e.g., by invoking a function in the caller.
Re-entrancy is one of the most common culprits behind vulnerabilities, and it was also exploited in the infamous ``The DAO'' attack~\cite{finley2016million}.
In Section~\ref{sec:lessvulnerable}, we discuss how the model behind VeriSolid prevents re-entrancy.

\subsubsection{``Denial of Service''~\cite{atzei2017survey}}
If a function involves sending ether using \texttt{transfer} or making a high-level function call to another contract, then the recipient contract can ``block'' the execution of this function by always throwing an exception.
Such vulnerabilities can be detected with VeriSolid using a type of liveness properties (see Section~\ref{sec:templates}), as we do for ``King of Ether 2'' (see Section~\ref{sec:kingOfTheEtherVerification}).

\subsubsection{Deadlocks}
A contract may end up in a ``deadlock'' state (either accidentally or through adversarial action), in which it is no longer possible to withdraw or transfer currency from the contract.
This means that the currency stored in the contract is practically lost, similar to what happened to the Parity multisignature wallet contracts~\cite{newman2017security}.
VeriSolid can verify if a contract model is deadlock-free, without requiring the developer to specify any property (Section~\ref{sec:verificationProcess}).

\subsection{Modeling and Verification with BIP and nuXmv}
\label{app:BIPmodeling}

We recall the necessary concepts of the Behavior-Interaction-Priority (BIP) component framework~\cite{basu2011rigorous}. BIP has been used for constructing several correct-by-design systems, such as robotic systems and satellite on-board software~\cite{said2014model,mavridou2016satellite,basu2008incremental}. Systems are modeled in BIP by superposing three layers: Behavior, Interaction, and Priority. The \textit{behavior} layer consists of a set of components represented by transition systems. Each component transition is labeled by a \textit{port}, which specifies the transition's unique name. Ports form the interface of a component and are used for interaction with other components. 
Additionally, each transition may be associated with a set of \textit{guards} and a set of \textit{actions}. A guard is a predicate on variables that must be true to allow the execution of the associated transition. An action is a computation triggered by the execution of the associated transition. Component interaction is described in the \textit{interaction} and \textit{priority} layers. 
We omit the explanation of these two layers, which are not used in this paper.

In order to check behavioral correctness of a system under design, formal verification is essential. While alternative approaches, such as simulation and testing, rely on the selection of appropriate test input for an adequate coverage of the program's control flow, formal verification (e.g., by model checking) guarantees full coverage of execution paths for all possible inputs. 
Thus, it provides a \textit{rigorous} way to assert (or deny) that a system model meets a set of properties.


In VeriSolid, we verify deadlock-freedom using the state space exploration analysis provided by BIP. 
This analysis checks deadlock by default, as it is an essential correctness property.
For the verification of safety and liveness properties, we use the BIP-to-NuSMV tool\footnote{%
  \url{http://risd.epfl.ch/bip2nusmv}
} to translate our BIP models into NuSMV,  the input language of the nuXmv symbolic model checker~\cite{bliudze2015formal}. The developer must give as input the properties to be verified directly as temporal logic formulas or by using natural language templates provided by our tool. 
The template input is used to  generate (\emph{Computation Tree Logic}) CTL specifications which are checked by the nuXmv tool.
  If a property is violated, the user gets a counterexample transition sequence that exemplifies the violation. Counterexamples help the user to locate the error back to the input model and identify its cause. The correctness of the BIP-to-NuSMV transformation based on bi-simulation was proved by Noureddine et al.~\cite{noureddine2014reduction}.


\section{Augmentation Algorithms and Equivalence Proof}
\label{app:generalAugmentation}

\subsection{Conformance Transformation}
\label{app:conformance}

First, we introduce Algorithm~\ref{alg:Conformance} for replacing the fallback and initial  actions, which model the fallback function and constructor of a Solidity contract, with functionally equivalent transitions.
Since the fallback function may be called in any state, the algorithm adds to each state a transition that does not change the state and whose action is the fallback action.
Then, the algorithm adds a new initial state and a transition from the new to the original initial state, whose action is the initial action.

\newcommand{\addTransAct}[3]{\textbf{add} transition from $#1$ to $#2$ with action $#3$}
\newcommand{\addTransGrd}[3]{\textbf{add} transition from $#1$ to $#2$ with guard $#3$}

\begin{algorithm}
\caption{\textit{Conformance}$(D, S,$ $S_F,$ $s_0,$ $a_0, a_F, V, T)$}
\label{alg:Conformance}
\begin{algorithmic}[1]
\Statex{\textbf{Input:} model $(D, S,$ $S_F,$ $s_0,$ $a_0, a_F, V, T)$}
\Statex{\textbf{Result:} model $(D, S,$ $S_F,$ $s_0,$ $V, T)$}
\For{state $s \in S$}
  \State{\addTransAct{s}{s}{a_F}}
\EndFor
\State{\textbf{add} state $s_I$}
\State{\addTransAct{s_I}{s_0}{a_I}}
\State{\textbf{change} initial state $s_0 := s_I$}
\end{algorithmic}
\end{algorithm}

\subsection{Augmentation Transformation}
\label{sec:augmentation}

Next, we introduce algorithms for translating a model with compound, selection, loop, etc. statements into a model with only variable declaration and expression statements. 
We first describe Algorithm~\ref{alg:AugmentStatement}, which translates a single transition with an arbitrary statement into a set of states and internal transitions with only variable declaration, expression, and return statements.
Then, we describe Algorithm~\ref{alg:AugmentModel}, which translates an entire model with the help of Algorithm~\ref{alg:AugmentStatement}.
Our augmentation algorithms are based on the small-step operational semantics of our supported Solidity Statements provided in Appendix~\ref{app:soliditySemantics}.

\begin{algorithm}
\caption{\textit{AugmentStatement}$(a, s_o, s_d, s_r)$}
\label{alg:AugmentStatement}
\begin{algorithmic}[1]
\Statex{\textbf{Input:} statement $a$, origin state $s_o$, destination state $s_d$, return state $s_r$}
\If{$a$ is variable declaration statement $\vee$ $a$ is event statement $\vee$ $a$ is expression statement}
  \State{\addTransAct{s_o}{s_d}{a}} 
\ElsIf{$a$ is \texttt{return} statement}
  \State{\addTransAct{s_o}{s_r}{a}} 
\ElsIf{$a$ is compound statement $\texttt{\{} a_1\texttt{;}\,a_2\texttt{;}\,\ldots\texttt{;}\,a_N\texttt{\}}$}
  \For{$i = 1, 2, \ldots, N-1$}
    \State{\textbf{add} state $s_i$}
  \EndFor
  \State{\textit{AugmentStatement$(a_1, s_o, s_1, s_r)$}}
  \For{$i = 2, 3, \ldots, N-1$}
    \State{\textit{AugmentStatement$(a_i, s_{i - 1}, s_i, s_r)$}}
  \EndFor
  \State{\textit{AugmentStatement$(a_N, s_{N-1}, s_d, s_r)$}}
\ElsIf{$a$ is selection statement \texttt{if\,($c$)\,$a_T$\,\,else\,\,$a_F$}}
  \State{\textbf{add} state $s_T$}
  \State{\addTransGrd{s_o}{s_T}{c}}
  \State{\textit{AugmentStatement$(a_T, s_T, s_d, s_r)$}}
  \State{\textbf{add} state $s_F$}
  \State{\addTransGrd{s_o}{s_F}{!(c)}}
  \State{\textit{AugmentStatement$(a_F, s_F, s_d, s_r)$}}
\ElsIf{$a$ is selection statement \texttt{if\,($c$)\,$a_T$}}
  \State{\textbf{add} state $s_T$}
  \State{\addTransGrd{s_o}{s_T}{c}}
  \State{\textit{AugmentStatement$(a_T, s_T, s_d, s_r)$}}
  \State{\addTransGrd{s_o}{s_d}{!(c)}}
\ElsIf{$a$ is loop statement \texttt{for\,($a_I$;\,$c$;\,$a_A$)\,$a_B$}}
  \State{\textbf{add} states $s_I, s_C, s_B$}
  \State{\textit{AugmentStatement$(a_I, s_o, s_I, s_r)$}}
  \State{\addTransGrd{s_I}{s_d}{!(c)}}
  \State{\addTransGrd{s_I}{s_C}{c}}
  \State{\textit{AugmentStatement$(a_B, s_C, s_B, s_r)$}}
  \State{\textit{AugmentStatement$(a_A, s_B, s_I, s_r)$}}
\ElsIf{$a$ is loop statement \texttt{while\,($c$)\,$a_B$}}
  \State{\textbf{add} state $s_L$}
  \State{\addTransGrd{s_o}{s_d}{!(c)}}
  \State{\addTransGrd{s_o}{s_L}{c}}
  \State{\textit{AugmentStatement$(a_B, s_L, s_o, s_r)$}}
\EndIf
\end{algorithmic}
\end{algorithm}

Algorithm~\ref{alg:AugmentStatement}, called \textit{AugmentStatement}, takes as input a Solidity statement, an origin, destination, and return state, and it creates a set of states and transitions that implement the input statement using only variable declaration, expression, and return statements as actions.
Note that before invoking this algorithm, Algorithm~\ref{alg:AugmentModel} removes the original transition between the origin and destination states; hence, this algorithm creates all transitions (and states) from scratch.
If the statement is a variable declaration, event, or expression statement, then the algorithm simply creates a transition from the origin to the destination state without any guards and having the statement as an action.
If the statement is a return statement, then it creates a transition from the origin to the return state.
Note that the return state is preserved by all recursive calls to \textit{AugmentStatement}, and it is initialized with the destination of the original transition by Algorithm~\ref{alg:AugmentModel}.

If the statement is a compound, selection, or loop statement, Algorithm~\ref{alg:AugmentStatement} creates a set of states and transitions.
For a \emph{compound statement} (i.e., list of statements), the algorithm creates a set of new states, each of which corresponds to the execution stage after an inner statement (except for the last one), and it invokes itself (i.e., \textit{AugmentStatement}) for each inner statement. 
For a \emph{selection statement} with an \texttt{else} (i.e., false) branch, it creates two states, which correspond to the true and false branches.
Then, 
it creates transitions to these states with the branch condition and its negation as guards, and invokes itself for both the true and false body statements. 
If the selection statement does not have an \texttt{else} branch, then the false branch is replaced by a simple transition 
to the destination state with the negation of the condition as a guard. 
Finally, given a \texttt{for} \emph{loop statement}, it creates three states, which model three stages of the loop execution: after initialization, after each time the loop condition is evaluated to true, and after each execution of the body.
Then, it invokes itself with the initialization statement, creates transitions with the loop condition and its negation (leading to the second state or the destination state), and then completes the loop by invoking itself for the body and afterthought statements.
For a \texttt{while} loop statement, it needs to create only one new state since there is no initialization or afterthought statement.


\begin{algorithm}
\caption{\textit{AugmentModel}$(D, S,$ $S_F,$ $s_0,$ $V, T)$}
\label{alg:AugmentModel}
\begin{algorithmic}[1]
\Statex{\textbf{Input:} model $(D, S,$ $S_F,$ $s_0,$ $V, T)$}
\Statex{\textbf{Result:} model $(D, S,$ $S_F,$ $s_0,$ $V, T)$}
\For{transition $t \in T$}
  \State{\textbf{remove} transition $t$}
  \State{\textbf{add} state $s_{grd}$}
  \State{\addTransGrd{t^{from}}{s_{grd}}{g_t}}
  \If{action $a_t$ cannot raise exception}
    \State{\textit{AugmentStatement}$(a_t, s_{grd}, t^{to}, t^{to})$}
  \Else
    \State{\addTransGrd{s_{grd}}{t^{from}}{\textnormal{``revert''}}}
    \State{\textbf{add} state $s_{rvrt}$}
    \State{\addTransGrd{s_{grd}}{s_{rvrt}}{\textnormal{``!revert''}}}
    \State{\textit{AugmentStatement}$(a_t, s_{rvrt}, t^{to}, t^{to})$}
  \EndIf
\EndFor
\end{algorithmic}
\end{algorithm}

Algorithm~\ref{alg:AugmentModel}, called \textit{AugmentModel}, takes as input a model that can have any set of supported statements as actions, and it translates the model into one that has only variable declaration, expression, and return statements.
It does so by iterating over the transitions and replacing each transition with a set of states and transitions using Algorithm~\ref{alg:AugmentStatement}.
Furthermore, it also augments the transition to consider the possibility that the transition is \emph{reverted} due to an exception (e.g., failure of a high-level function call or \texttt{\small transfer}). More specifically, for each original transition, it first removes the transition, then adds a state $s_{grd}$ and a transition from the origin to $s_{grd}$ with the original guard.
If the action contains a statement that can result in an exception, the algorithm also adds a state $s_{rvrt}$, a transition from state $s_{grd}$ to state $s_{rvrt}$, and a transition from state $s_{grd}$ to the origin state.
During verification, our tool considers the possibility of the entire transition being reverted using this branch.
Finally, the algorithm invokes \textit{AugmentStatement} with the original action and original destination.

\subsection{Observational Equivalence Proof}
\label{app:correctnessProof}

Below we provide the proof of Theorem~\ref{th:bisim}.

\begin{proof}
We are going to prove that all three conditions hold for some pair ($q, r$), for which certain criteria hold.

Before that, let us repeat a set of preliminary assumptions for the states and transitions in both systems.
From the transformation algorithm, it holds that for each state $q$, there is exactly one corresponding state $c(q) \in S_E$, at which there can be invoked exactly the same functions as at $q$. 

The execution semantics of a function says that $\alpha$ may be reverted , or that it may be executed normally (finished). There are $\alpha^{fin} , \alpha^{rev} \in A$ transitions for representing each of these cases. For each such $\alpha$ in the transitions of $q$, there is a set of outgoing paths $P_a$ at $c(q)$, where both $\alpha$ and $P_a$ represent the same execution semantics, only that paths consist of distinct transitions for each Solidity code statement in $\alpha$ (branching in paths is caused due to \textit{if} and \textit{while} constructs). 
Each $P_a$ can be represented by the regular expression $\beta^{call}\beta^{\star}\alpha$, where $\beta^{call}$ is the function call, each $\beta$-transition is an arbitrary code statement, and $\alpha$ is either $\alpha^{rev}$ or $\alpha^{fin}$. 

\figurename~\ref{img:proofImage} shows a state $q \in S_I$ (bottom) with two transitions $\alpha^{rev}$ and $\alpha^{fin}$ and its corresponding $r=c(q) \in S_E$ (top) with the outgoing $P_{\alpha^{rev}}$ and $P_{\alpha^{fin}}$. We will prove the relationship $R$ denoted by the dotted lines, i.e., that $(q, r), (q, r1), (q, r2), (q, 3), (q', r') \in R$ for each such $\alpha \in A$. In other words, if $r$ is correspondent to $q$, then it is equivalent with $q$ and all the other $r_i$ that are reachable in the path, are also equivalent with $q$, except for $r'$, which is equivalent with $q'$.
If we prove this for one $\langle q, r, \alpha \rangle$ tuple, then it holds for all of them.

\begin{figure}[h]
\center
\includegraphics[width=.4\textwidth]{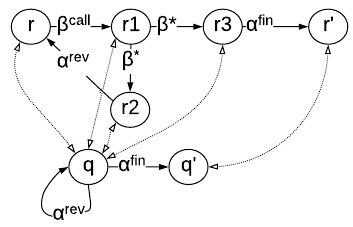} 
\label{img:proofImage}
\caption{Abstract representation of states in the smart contract (bottom) and the augmented system (top) ($R$ is shown with dotted lines).}
\end{figure}

First, let us prove that $(q, r)\in R$. For each $\alpha^{fin}\in A$, such that $q \xrightarrow{\alpha^{fin}}q'$ there is a $P_{\alpha^{fin}}$, such that $r \xrightarrow{P_{\alpha^{fin}}}r'$ and $P_{\alpha^{fin}}=\beta*\alpha^{fin}$, where $\beta \in B$ and $\alpha^{fin} \in A$. 
Moreover, $q'$ and $r'$  are corresponding states just like $q$ and $r$, thus, if $(q, r)\in R$ is proved, so is $(q', r')\in R$.
As with  each $\alpha^{fin}\in A$, also for each $\alpha^{rev}\in A$ there is a $P_{\alpha^{rev}}$, such that $r \xrightarrow{P_{\alpha^{rev}}}r$, where $\beta \in B$ and $\alpha^{rev} \in A$. Moreover, the final states being $q$ and $r$  are now being  proved equivalent. So, far we have proved Property 1 for $(q, r)$. Property 2 does not apply since there are no transitions of $A$ starting from $r$. 
For Property 3, we have to prove that $(q,r1)\in R$, since $r1$ is the only state that is reachable from $r$ through transitions of $B$.  We will prove $(q,r1)\in R$ at a later step.
Since the three Properties hold, $(q, r)\in R$ has been proved. Note that since $(q, r)\in R$, it follows that $(q', r')\in R$.

Let us prove now that $(q, r1)\in R$. For each $\alpha^{fin}\in A$, such that $q \xrightarrow{\alpha^{fin}}q'$ there is a $P_{\alpha^{fin}}$, such that $r1 \xrightarrow{P_{\alpha^{fin}}}r'$ and $P_{\alpha^{fin}}=\beta*\alpha^{fin}$, where $\beta \in B$ and $\alpha^{fin} \in A$. Moreover, it has been proved that $(q', r')\in R$. Similarly, for each $\alpha^{rev}\in A$ there is a $P_{\alpha^{rev}}$, such that $r1 \xrightarrow{P_{\alpha^{rev}}}r$, where $\beta \in B$, $\alpha^{rev} \in A$ and the final states $q$ and $r$  are equivalent. Property 2 does not apply. For Property 3, we have to prove that $(q,r2)\in R$ and $(q,r3)\in R$, since $r2$ and $r3$ are the only states that are reachable from $r1$ through transitions of $B$.  We will prove $(q,r2)\in R$ and $(q,r3)\in R$ at a later step.
Since the three Properties hold, $(q, r1)\in R$ has been proved. 

Let us prove that $(q, r2)\in R$ and that $(q, r3)\in R$. 
For each $\alpha^{fin}\in A$, such that $q \xrightarrow{\alpha^{fin}}q'$ there is a $\alpha^{fin}$, such that $r3 \xrightarrow{P_{\alpha^{fin}}}r'$ and $P_{\alpha^{fin}}\alpha^{fin}$, where $\beta \in B$ and $\alpha^{fin} \in A$. Moreover, it has been proved that $(q', r')\in R$. 
Similarly, for each $\alpha^{rev}\in A$ there is a $P_\alpha^{rev}$, such that $r2 \xrightarrow{\alpha^{rev}}r$, where $\beta \in B$, $\alpha^{rev} \in A$ and the final states $q$ and $r$  are equivalent. 
Property 2 holds since for each $\alpha^{fin}\in A$ and  $\alpha^{rev}\in A$, such that $r3 \xrightarrow{\alpha^{fin}}r'$ and $r2 \xrightarrow{\alpha^{rev}}r$, there is an $\alpha^{fin}\in A$ (resp. $\alpha^{rev}\in A$) such that $q \xrightarrow{\alpha^{fin}}q'$ (resp. $q \xrightarrow{\alpha^{fin}}q$) and it has been proved that $(q',r')\in R$ (resp. 
$(q,r)\in R$).
Property 3 does not apply, since there are no states that are reachable from $r2$ or $r3$ through transitions of $B$. Since the three Properties hold, $(q, r2)\in R$ and $(q, r3)\in R$ have been proved.

\end{proof}

\section{Solidity Code Generation}
\label{sec:transformation}

The VeriSolid code generator is an extension of the FSolidM code generator~\cite{mavridou2018designing}. The code generation takes as input the initial transition system modeled by the developer. To generate  Solidity code, it follows directly the operational semantics of the transition system defined in Appendix~\ref{app:transitionSemantics}.
We first provide an overview of the key differences between the two generators, and then present the VeriSolid code generator. 
We refer the reader to~\cite{mavridou2018designing} for a detailed presentation of the FSolidM code generator.

Compared to FSolidM, the VeriSolid generator contains the following main differences:
\begin{itemize}
\item At the beginning of each transition, the value of the state variable \texttt{state} is set to \texttt{InTransition} (if the transition has a non-empty action).
\item A constructor is generated from the initial action $a_0$.
\item A fallback function is generated from the fallback action~$a_F$.
\item To maintain functional equivalence between the model and the generated code, FSolidM code-generator plugins (see~\cite{mavridou2018designing}) are not supported.
\end{itemize}

The input of the VeriSolid code generator is a smart contract that is defined (see Definition~\ref{def:smartContract}) as a transition system $(D, S, S_F, s_0, a_0,$ $a_F, V, T)$.
In addition, the developer specifies the $name$ of the contract.
Further, for each transition $t \in T$, the developer specifies $t^{payable}$, which is true if the function implementing transition $t$ should be payable and false otherwise.

For each contract variable or input variables (i.e., function argument) $v \in \mathbb{I} \times \mathbb{T}$, we let $name(v) \in \mathbb{I}$ and $type(v) \in \mathbb{T}$ denote the name and type of the variable, respectively.
We use $\code{fixed-width}$ font for generated code, and we use and $\lbl{italic}$ font for elements that are replaced with input or specified later.

\begin{align*}
\lbl{Contract} ::= ~&\code{contract } name \code{ \{} \\
& \ind \lbl{StatesDefinition} \\
& \ind \lbl{VariablesDefinition} \\ 
& \ind \lbl{Constructor} \\
& \ind \lbl{Fallback} \\
& \ind \lbl{Transition}(t_1) \\
& \ind \ldots \\
& \ind \lbl{Transition}(t_{|T|}) \\
& \code{\}}
\end{align*}
where $\{ t_1, \ldots, t_{T}\}$ is the set of transitions $T$.

\begin{align*}
\lbl{StatesDefinition} ::= ~&\code{enum States \{} \\
& \ind \code{InTransition, } s_0 \code{,} \ldots \code{,} s_{|S| - 1} \\
& \code{\}} \\
& \code{States private state;}
\end{align*}
where $\{ s_0, \ldots, s_{|S| - 1} \}$ is the set of states $S$.

\begin{align*}
\lbl{VariablesDefinition} ::= ~& D \\
& type(v_1) ~ name(v_1) \code{;} \\
& \ldots \\
& type(v_{|V|}) ~ name(v_{|V|}) \code{;}  \\
& \code{uint private creationTime = now;}
\end{align*}
where $D$ is the set of custom event and type definitions, and $\{ v_1, \ldots,$ $v_{|V|} \}$ is the set of contract variables $V$.

\begin{align*}
\lbl{Constructor} ::= ~&\code{constructor () public \{ } \\
& \ind \lbl{Action}(a_0, \code{States.}s_0) \\
& \ind \code{state = States.}s_0\code{;} \\
& \code{\}} 
\end{align*}
where $s_0$ and $a_0$ are the initial state and action, respectively.

\begin{align*}
\lbl{Fallback} ::= ~&\code{function () payable public \{} \\ 
& \ind \code{State memory currentState = state;} \\
& \ind \lbl{Action}(a_F, \code{currentState}) \\
& \ind \code{state = currentState;} \\
& \code{\}} 
\end{align*}
where $a_F$ is the fallback action.

\begin{align*}
\lbl{Transition}(t) ::= ~&\code{function } t^{name} 
\code{(} type(i_1)~ name(i_1)\code{, }\\
& \ind \ind \ldots  \code{, }  type\left(i_{|t^{input}|}\right)~ name\left(i_{|t^{input}|}\right) \code{)} \\ 
& \ind \ind \code{public } \lbl{Payable}(t) ~ \lbl{Returns}(t) \code{ \{} \\
& \ind \code{require(state == States.} t^{from} \code{);} \\
& \ind \code{require (} g_t \code{);} \\
& \ind \lbl{Action}(a_t, \code{States.}t^{to}) \\
& \ind \code{state = States.}t^{to}\code{;} \\
& \code{\}} 
\end{align*}
where $t^{name}$ is the name of transition $t$, $\left\{i_1, \ldots, i_{|t^{input}|}\right\}$ is the set of parameter variables (i.e., arguments) $t^{input}$,
$g_t$ and $a_t$ are the guard and action, and $t^{from}$ and $t^{to}$ are the source and destination states. 

If $t^{payable}$ is true, then $\lbl{Payable}(t) ::= \code{payable}$; otherwise, $\lbl{Payable}(t)$ is empty.
If return type is $t^{output} = \emptyset$, then $\lbl{Returns}(t)$ is empty. Otherwise, it is 
$\lbl{Returns}(t) ::=  \code{returns (} t^{output} \code{)} $ 

If $a = \emptyset$ (i.e., empty action statement), then $\lbl{Action}(a, s)$ is empty.
Otherwise, 
\begin{align*}
\lbl{Action}(a, s) ::= ~ & \code{state = States.InTransition;} \\
& \lbl{SafeAction}(a, s)
\end{align*}

Finally, $\lbl{SafeAction}(a, s)$ simply means $a$, but replacing any $$\code{return } \lbl{expression};$$
or
$$\code{return;}$$ statement with a $$\code{\{ state = } s \code{; return } \lbl{expression} \code{; \}}$$
or
$$\code{\{ state = } s \code{; return; \}}$$
compound statement in $a$. Note that this applies to all inner statements within $a$ (body statements within selection statements, loop statements, etc.).

\section{Blind Auction}
\label{sec:completeBlindAuction}

\subsection{Complete Augmented Model}
\label{app:completeBlindAuctionAugmented}
\begin{figure}[h!] 
\centering
\includegraphics[width=\textwidth]{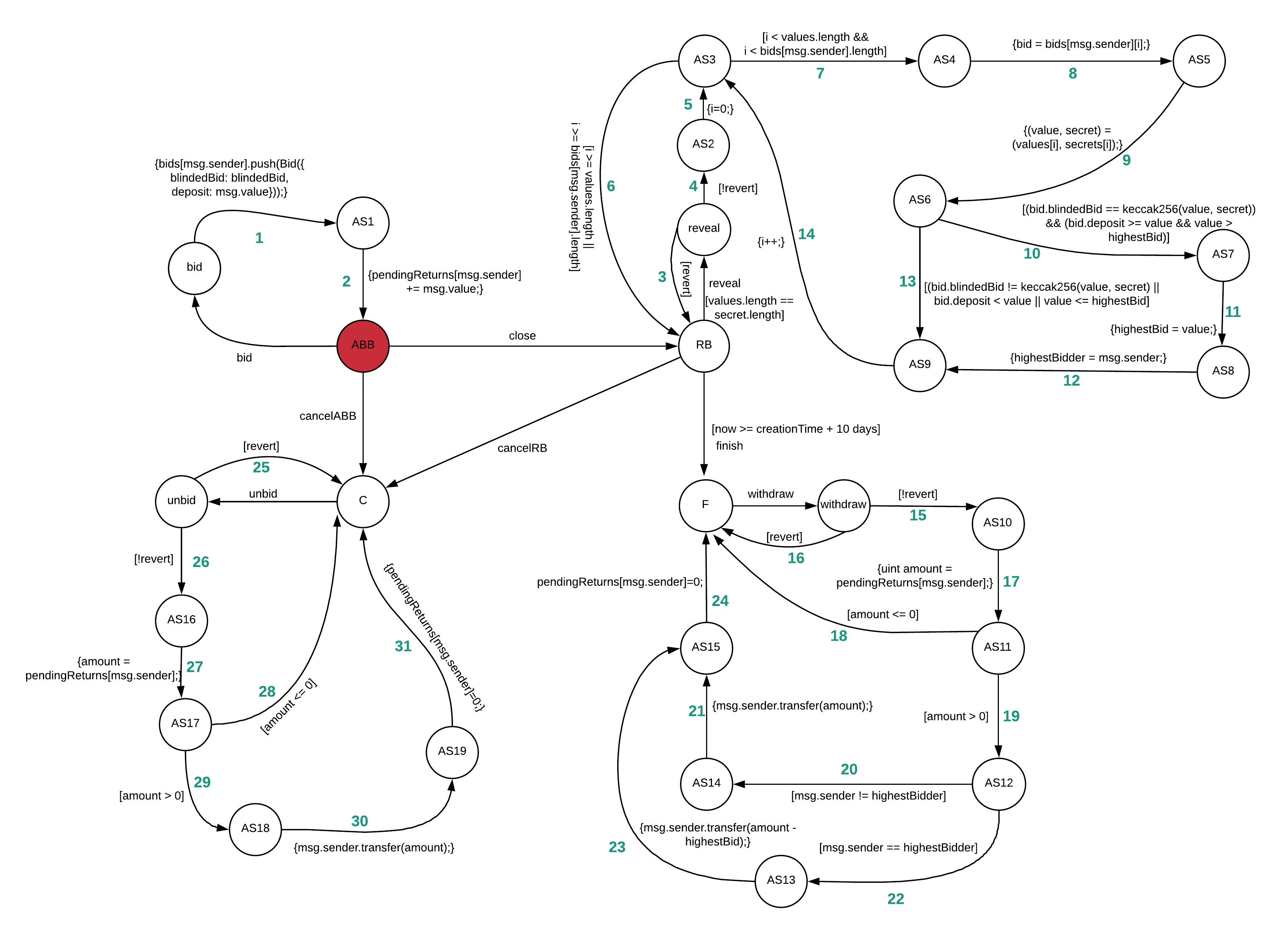}
\caption{Augmented model of the Blind Auction. 
}
\label{fig:completeBlind}
\end{figure}

Figure \ref{fig:completeBlind} presents the complete augmented model of the Blind Auction Contract.

\subsection{Solidity Code}
Below we present the Solidity code generated from VeriSolid.

\begin{lstlisting}[style=customjava]
contract BlindAuction{
    //States definition
    enum States {
    	InTransition,
        ABB,
        RB,
        F,
        C
    }
    States private state = States.ABB;
    
    //Variables definition
    struct Bid {
        bytes32 blindedBid;
        uint deposit;
    }
    mapping(address => Bid[]) private bids;
    mapping(address => uint) private pendingReturns;
    address private highestBidder;
    uint private highestBid;
    uint private creationTime = now;
    
    //Transitions
    
    //Transition bid
    function bid (bytes32 blindedBid) public payable
    {
        require(state == States.ABB);
        //State change
        state = States.InTransition;
        //Actions
        bids[msg.sender].push(Bid({
            blindedBid: blindedBid,
            deposit: msg.value
        }));
        pendingReturns[msg.sender] += msg.value;
        //State change
        states = States.ABB;
    }
    
    //Transition close
    function close() public
    {
        require(state == States.ABB);
        //Guards
        require(now >= creationTime + 5 days);
        //State change
        state = States.RB;
    }
    
    //Transition reveal
    function reveal(uint[] values, bytes32[] secrets) public
    {
        require(state == States.RB);
        //Guards
        require(values.length == secrets.length);
        //State change
        state = States.InTransition;
        //Actions
        for (uint i = 0; i < values.length &&
                i < bids[msg.sender].length; i++) {
            var bid = bids[msg.sender][i];
            var (value, secret) = (values[i], secrets[i]);
            if (bid.blindedBid == keccak256(value, secret) && 
                    bid.deposit >= value && 
                    value > highestBid) {
                highestBid = value;
                highestBidder = msg.sender;
            }
        }
        //State change
        state = States.RB;
    }
    
    //Transition finish
    function finish() public
    {
        require(state == States.RB);
        //Guards
        require(now >= creationTime + 10 days);
        //State change
        state = States.F;
    }
    
    //Transition cancelABB
    function cancelABB() public
    {
        require(state == States.ABB);
        //State change
        state = States.C;
    }
    
    //Transition cancelRB
    function cancelRB() public
    {
        require(state == States.RB);
        //State change
        state = States.C;
    }
    
    //Transition withdraw
    function withdraw() public
    { 
        require(state == States.F);
        //State change
        state = States.InTransition;
        // Actions
        uint amount = pendingReturns[msg.sender];
        if (amount > 0) {
            if (msg.sender!= highestBidder)
                msg.sender.transfer(amount);
            else
                msg.sender.transfer(amount - highestBid);
            pendingReturns[msg.sender] = 0;
        }
        //State change
        state = States.F;
    }
    
    //Transition unbid
    function unbid() public
    {
        require(state == States.C);
        //State change
        state = States.InTransition;
        //Actions
        uint amount = pendingReturns[msg.sender];
        if (amount > 0) {
            msg.sender.transfer(amount);
            pendingReturns[msg.sender] = 0;
        }
        //State change
        state = States.C;
    }   
}
\end{lstlisting}

\clearpage
\section{Further Example Models}

\subsection{DAO Model}
\label{sec:dao}

\begin{figure*} 
\centering
\includegraphics[width=\textwidth]{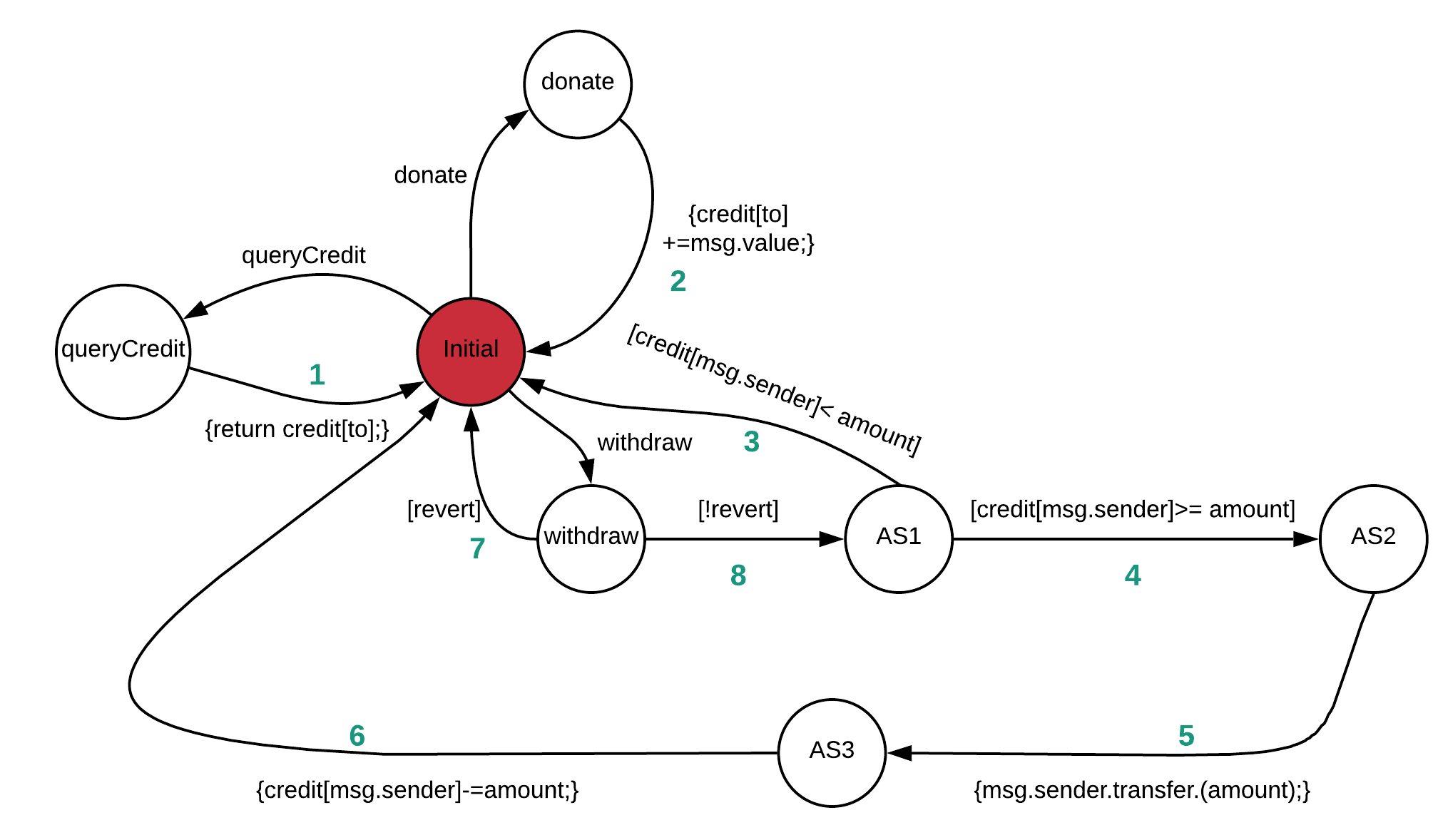}
\caption{Simplified model of the DAO contract.
}
\label{fig:dao}
\end{figure*}

The DAO contracts implemented a crowd-funding platform, which raised approximately \$150 million before being attacked in June 2016. Here, we present a simplified version of the DAO contract, which allows participants to \texttt{donate} ether to fund contracts, while contracts can then \texttt{withdraw} their funds. The augmented model of the contract is presented in Figure~\ref{fig:dao}.

By verifying the safety property presented in Table 1, we can guarantee that none of the two attacks presented in~\cite{atzei2017survey} can be successful on our contract. Both of these attacks are possible if the contract sends the amount of ether before decreasing the credit and in the meantime an attacker makes another function call, e.g., to \texttt{withdraw}. Although the former is true for our transition system, i.e., transition $6$ happens after transition $5$, by-design our contract changes state when the \texttt{withdraw} function is called. In particular, our contract goes from the \texttt{Initial} state to the \texttt{withdraw} state and thus, after executing transition $5$, the attacker cannot make another function call. In other words, $6$ will always happen right after the execution of $5$.

\subsection{The King Of the Ether Throne Models}
\label{sec:king}

\begin{figure*} 
\centering
\includegraphics[width=\textwidth]{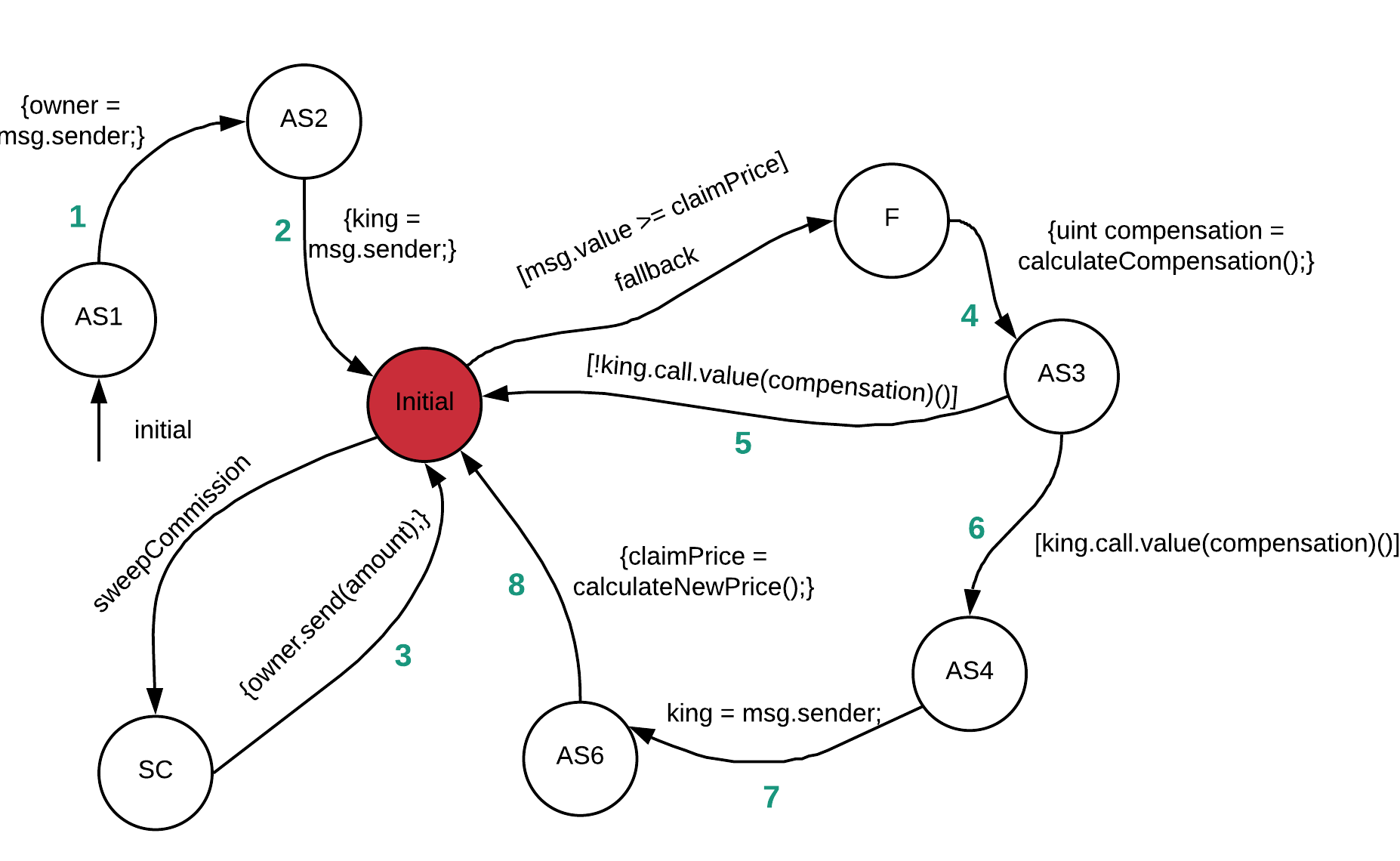}
\caption{King of Ether 1. 
}
\label{fig:king1}
\end{figure*}

\begin{figure*} 
\centering
\includegraphics[width=\textwidth]{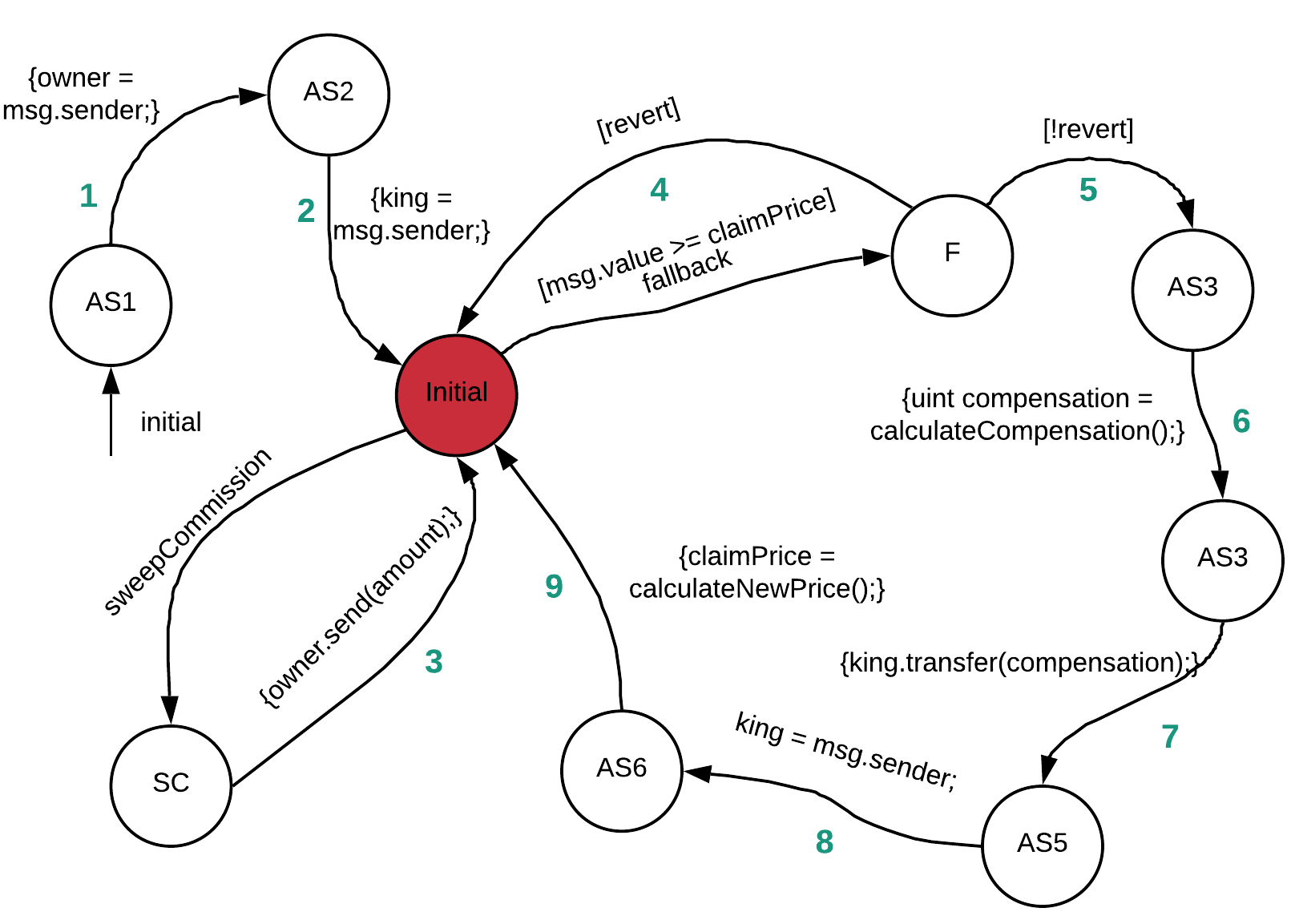}
\caption{King of Ether 2. 
}
\label{fig:king2}
\end{figure*}

The ``King of the Ether Throne'' is a game where players compete for acquiring the title of the King. If someone wishes to be the king, he must pay an amount of ether (which increases monotonically) to the current king. In Figures~\ref{fig:king1} and~\ref{fig:king2}, we present the models of two versions of the King of the Ether Throne contract~\cite{atzei2018formal}. 

The denial of service vulnerability can be exploited in these contracts. To see why, consider an attacker Mallory, whose \texttt{fallback} just throws an exception. The adversary sends the right amount of ether, so that Mallory becomes the new king. Now, nobody else can get her crown, since every time the King of the Either Throne contract (either of the two versions) tries to send the compensation to Mallory, her fallback throws an exception, preventing the coronation to succeed. In particular, ``King of Ether 1'' uses \texttt{call} which is going to return \texttt{false}, while ``King of Ether 2'' uses \texttt{transfer} that is going to be \texttt{reverted}. We were able to check that our models have this denial of service vulnerability by model checking the liveness properties presented in Table~1.

\subsection{Resource Allocation Contract}
\label{app:transax}

 
TRANSAX is a blockchain-based platform for trading energy futures~\cite{laszka2018transax}.
The core of this platform is a smart contract that allows energy producers and consumers to post offers for selling and buying energy.
Since optimally matching selling offers with buying offers can be very expensive computationally, the contract relies on external solvers to compute and submit solutions to the matching problem, which are then checked by the contract.
 
We defined a set of safety properties for this contract
(Table~\ref{tab:transaxresults} presents a subset of these properties). We were able to find a bug in the action of the \texttt{finalize} transition:

\begin{lstlisting}[style=customjava] 
// action of finalize transition
if (solutions.length > 0) {
    Solution storage solution = solutions[bestSolution];
    for (uint64 i = 0; i < solution.numTrades; i++) {
        Trade memory trade = solution.trades[i];
        emit TradeFinalized(trade.sellingOfferID, 
        trade.buyingOfferID, trade.power, trade.price);
    }  
    solutions.length = 0;
    offers.length = 0;
} 
// offers.length = 0; SHOULD HAVE BEEN HERE
cycle += 1;
\end{lstlisting}
This bug was immediately detected as a violation of our first safety property shown in Table~\ref{tab:transaxresults}.

\begin{table}[t]
  \caption{Analyzed properties and verification results for the Resource Allocation case study.}
  \centering
  \begin{tabular}{ p{2.2cm} p{6.8cm} p{1.3cm} p{1.1cm} }
  
    \textbf{Case Study} & \textbf{Properties} & \textbf{Type}  & \textbf{Result} \\ \hline  \hline
    \vspace{-.8cm}
    \multirow{1}{2cm}{
    \begin{tabular}{p{2cm}}
        Resource\\ Allocation\\
        states: 3487
    \end{tabular}} & 
    \hspace{.1cm}\begin{tabular}{p{6cm}}
    (i) if \texttt{close} happens, \texttt{postSellingOffer} or \texttt{postBuyingOffer} can happen only after \texttt{finalize.offers.length=0}
    \end{tabular} & Safety & Violated \\
  
 &  \hspace{.1cm}\begin{tabular}{p{6cm}} 
 (ii) \texttt{register.prosumers[msg.sender]= prosumerID} cannot happen after \texttt{setup} \end{tabular} 
  & \vspace{-.05cm} Safety & \vspace{-.05cm}Verified \\
  
  &  \hspace{.1cm}\begin{tabular}{p{6cm}} 
  (iii) \texttt{register} cannot happen after \texttt{setup} \end{tabular} & Safety & Verified \\
  
  &  \hspace{.1cm}\begin{tabular}{p{6cm}} 
  (iv) if \texttt{finalize} happens  \texttt{createSolution} or \texttt{addTrade}  can happen only after \texttt{close} \end{tabular} & Safety & Verified \\ \hline

  \end{tabular}
  \label{tab:transaxresults}
\end{table}

\newpage
\section{Extended Related Work}
\label{app:extended_related}

\subsubsection{{Vulnerability Types:}}
\label{sec:commonVuln} 

Motivated by the large number of smart-contract vulnerabilities, multiple research efforts investigate and establish taxonomies of common security vulnerabilities.
Atzei et al.\ provide a comprehensive taxonomy of  Ethereum smart-contract vulnerabilities, which identifies twelve common types~\cite{atzei2017survey}.
They show for nine of these types how an adversary can steal assets or inflict damage by exploiting a vulnerability.
In another effort,  
Luu et al.\ discuss four vulnerability types---which are also identified in~\cite{atzei2017survey}---and they propose various techniques for mitigating them~\cite{luu2016making}. 

\subsubsection{{Verification and Vulnerability Discovery}}
\label{sec:verification}

Both verification and vulnerability discovery are considered in the literature for identifying smart-contract vulnerabilities. 
The main advantage of our model-based approach is that it allows developers to specify desired properties with respect to a high-level model instead of, e.g., EVM bytecode, and also provides verification results and counterexamples in a developer-friendly, easy to understand, high-level form.
Further, our approach allows verifying whether a contract satisfies all desired security properties instead of detecting certain types of vulnerabilities; hence, it can detect atypical vulnerabilities.
Parizi et al. provide a survey and comparison of existing tools for automatic security testing of smart contracts~\cite{parizi2018empirical}.

For example, Hirai performs a formal verification of a smart contract that is used by the Ethereum Name Service~\cite{hirai2016formal}. 
However, this verification proves only one particular property and it involves relatively large amount of manual analysis.
In later work, Hirai defines the complete instruction set of the Ethereum Virtual Machine (EVM) in Lem, a language that can be compiled for interactive theorem provers~\cite{hirai2017defining}.
Using this definition, certain safety properties can be proven for existing contracts.
Atzei et al.\ propose a formal model of Bitcoin transactions, which enables formal reasoning, and they prove well-formedness properties of the Bitcoin blockchain~\cite{atzei2018formal}.
Bhargavan et al.\ outline a framework for verifying the safety and correctness of Ethereum contracts~\cite{bhargavan2016short}.
The framework is built on tools for translating Solidity and EVM bytecode contracts into $F^{*}$, a functional programming language aimed
at program verification.
Using the~$F^{*}$ representations, the framework can verify the correctness of the Solidity-to-bytecode compilation and detect certain vulnerable patterns. 
Tsankov et al. introduce a security analyzer for Ethereum contracts, called \textsc{Securify}~\cite{tsankov2018securify}.
To analyze a contract, \textsc{Securify} first symbolically encodes the dependence graph of
the contract in stratified Datalog~\cite{jeffrey1989principles}, and then it uses off-the-shelf Datalog solvers to check the satisfaction of properties, which can be described in a DSL.

Ellul and Pace use techniques from runtime
verification to build the \textsc{ContractLarva} tool, which enables extending contracts to detect violations at runtime and to offer monetary reparations in response to a violation~\cite{ellul2018runtime}.
Colombo et al. also argue that dynamic analysis can be used not only to detect errors but also to recover from them, and they discuss how to extend the \textsc{ContractLarva} tool to this end~\cite{colombo2018contracts}.

Luu et al.\ provide a tool called \textsc{Oyente}, which can analyze smart contracts and detect certain typical security vulnerabilities~\cite{luu2016making}.
They also recommend changes to the execution semantics of Ethereum, which would eliminate vulnerabilities of the four types that are discussed in their paper.
However, these changes would need to be adopted by all Ethereum clients. 
Building on \textsc{Oyente}, Albert et al. introduce the \textsc{EthIR} framework for analyzing Ethereum bytecode~\cite{albert2018ethir}.
\textsc{EthIR} can produce a rule-based representation of bytecode, which enables the application of existing analysis to infer properties of the EVM code.
Nikolic et al. present the \textsc{MAIAN} tool for detecting three types of vulnerable contracts, called prodigal, suicidal and greedy~\cite{nikolic2018finding}.
\textsc{MAIAN} allows detecting trace vulnerabilities (i.e., vulnerabilities across a sequence of invocations of a contract) by analyzing smart contract bytecode. According to their findings, more than $30$ thousand smart contracts deployed on the public Ethereum blockchain suffer from at least one vulnerability.
Fr{\"o}wis and B{\"o}hme define a heuristic indicator of control flow immutability to quantify the prevalence of contractual loopholes based on modifying the control flow of Ethereum contracts~\cite{frowis2017code}.
Based on an evaluation of all the contracts deployed on Ethereum, they find that two out of five contracts require trust in at least one third party.
%
Brent et al. introduce a security analysis framework for
Ethereum smart contracts, called \textsc{Vandal},
which converts EVM bytecode to semantic relations, which are then analyzed to detect vulnerabilities, which can be described in the Souffl\'e language~\cite{brent2018vandal}.
Mueller presents \textsc{Mythril}, a security analysis tool for Ethereum smart contracts with a symbolic execution backend, which can be used to detect vulnerabilities~\cite{mueller2018smashing}.
Stortz introduces \textsc{Rattle}, a static analysis framework for EVM bytecode that can recover control flow graph, lift it into SSA / infinite register form, and optimize it, facilitating further analyses~\cite{stortz2018rattle}.

%
%

\subsubsection{{Formal Operational Semantics}}
There are a number of research efforts that  focus on defining formal operational semantics for the EVM bytecode and the Solidity language. Hildenbrandt et al.~\cite{hildenbrandt2017kevm} formally define the semantics of EVM instructions in the K-framework~\cite{kframework} and validate them. In the same category falls the work of Grischchenko et al.~\cite{grishchenko2018semantic,grishchenko2018semantic2}, which presents a set of small-step semantics for the EVM bytecode. They formalized a large subset of their defined semantics in the F* proof assistant and validated them against the official Ethereum test suite. Additionally, they formally define
security properties for smart contracts, such as call integrity and atomicity. Both research efforts were able to find ambiguities in the official EVM specification~\cite{wood2014ethereum}.

Yang and Hang~\cite{yang2018lolisa} define big-step operational semantics for a large subset of the Solidity language. The work by Jiao et al.~\cite{jiao2018executable} defines small-step operational semantics for a subset of the Solidity language. Additionally, this work implements and validates the proposed semantics in the K-framework~\cite{kframework}. In our paper, we built on the small-step semantics defined in~\cite{jiao2018executable}, which enables us to reason about one computational step at a time. We extended their Solidity statement semantics to support exceptions and return values.

\subsubsection{{Design Patterns and Development}}
Bartoletti and Pompianu identify nine common design patterns in Ethereum smart contracts~\cite{bartoletti2017empirical}.
By studying the usage of patterns in publicly deployed contracts, they find that the most common one is a security pattern, called ``authorization,'' which is found in 61\% of all contracts. 
They also provide a taxonomy of Bitcoin and Ethereum contracts, dividing them into five categories based on their application domain, finding that the most common Ethereum contracts are financial and~notary. 
%
W\"ohrer and Zdun also study common design patterns in Ethereum smart contracts, based on Multivocal Literature Research~\cite{wohrer2018design}.
They provide a taxonomy consisting of 18 patterns, and study which patterns appear commonly and how these patterns map to Solidity coding practices.
%
O'Connnor introduces a typed, combinator-based, functional language, called Simplicity, for smart contracts~\cite{oconnor2017simplicity}. 
Simplicity is not Turing complete, which may limit its applicability, but also makes it amenable to static analysis.
%
Frantz and Nowostawski propose an approach for semi-automated translation of human-readable contract representations into computational
equivalents~\cite{frantz2016institutions}.
They also identify smart contract components that correspond to real-world institutions and propose a mapping; however, they do not provide formal guarantees or security assurances for the generated code.
%
Hu and Zhong propose a logic-based smart contract
model, called \textsc{Logic-SC}, based on semantics and syntax of Active-U-Datalog with temporal extensions; however, they do not consider security properties~\cite{hu2018method}.

\end{document}